\documentclass[onecolumn,showpacs,preprintnumbers,amsmath,amssymb,aps,pre]{revtex4}

\usepackage{epsfig}
\usepackage{psfrag}
\usepackage{subfigure}
\usepackage{amsmath}
\usepackage{amsmath}

\begin{document}

\title{Lattice Boltzmann simulations of phase separation in chemically reactive binary fluids}
\author{K.~Furtado}
\affiliation{Rudolf Peierls Centre for Theoretical Physics, 1 Keble Road,
Oxford, OX1 3NP, United Kingdom.}
\author{J.~M.~Yeomans}
\affiliation{Rudolf Peierls Centre for Theoretical Physics, 1 Keble Road,
Oxford, OX1 3NP, United Kingdom.}
\date{\today}

\begin{abstract}
We use a lattice Boltzmann method to study pattern formation in chemically reactive binary fluids in the regime where hydrodynamic effects are important. The coupled equations solved by the method are a Cahn-Hilliard equation, modified by the inclusion of a reactive source term, and the Navier-Stokes equations for conservation of mass and momentum. The coupling is two-fold, resulting from the advection of the order-parameter by the velocity field and the effect of fluid composition on pressure. We study the the evolution of the system following a critical quench for a linear and for a quadratic reaction source term. Comparison is made between the high and low viscosity regimes to identify the influence of hydrodynamic flows. In both cases hydrodynamics is found to influence the pathways available for domain growth and the eventual steady-states.
\end{abstract}

\maketitle
\section{Introduction}\label{Introduction}
The process of phase separation in chemically reactive mixtures has been considered by several authors. Glotzer \emph{et al} \cite{glotzer} and Christensen \emph{et al} \cite{christen} used a modification of the Cahn-Hilliard equation to investigate the effects of a linear reaction of the type $A\leftrightarrow B$ occurring simultaneously with phase separation following an instantaneous quench. In contrast to phase separation alone, domain coarsening was halted at a length-scale dependent on system parameters resulting in the `freezing in' of a spatially heterogeneous pattern. It was recognized that the steady-states resulted from competition between the demixing effects of phase separation and the equivalence of the chemical reaction term to an effective long-range repulsion \cite{christen,glotzer2}. Similar physics is seen in the phase ordering of block copolymers where an effective long-range interaction arises because of an interplay between interactions and steric constraints \cite{leiber,ohta}. In such systems pattern formation is a result of thermodynamic equilibrium. By contrast, in the systems we consider, the steady-states are maintained dynamically by the interplay of reaction and diffusion.

A number of chemically and structurally more complicated systems have been considered, numerically and theoretically, within the same framework of a modified Cahn-Hilliard equation. These include ternary mixtures \cite{okuz,liu,tong,balazs} and systems with orientational order \cite{reig}.

Here we investigate the effect of hydrodynamic interactions on phase ordering in a binary fluid mixture with chemical reactions using a lattice Boltzmann method. The case of the linear reaction has been considered before by Hou {\it et al} \cite{huo} by a different numerical method. We duplicate some of their results as a means of testing our approach and then consider the quadratic reaction mechanism $A + B \leftrightarrow 2B$. 

The inclusion of hydrodynamics is known to strongly affect the way in which an unreactive fluid mixture coarsens in the aftermath of a quench \cite{bray,siggia}. The growth exponent is found to increase from $\alpha=1/3$, for the purely diffusive case, to $\alpha=1$ or $\alpha=2/3$ for the viscous and inertial hydrodynamic regimes respectively. The new pathway for growth provided by hydrodynamics is transport of the bulk fluid down a pressure gradient established by variations in curvature \cite{siggia}. In two dimensions this minimises curvature by making domains circular, whereupon the effect vanishes and further coarsening can only occur by diffusion \cite{wagner}. In addition there is the possibility, investigated by Tanaka \cite{tan}, that the rapid decrease in interfacial area resulting from the hydrodynamic mechanism may leave the bulk phases unequilibrated and subject to a round of secondary phase separations. This suggests that coupling a modified Cahn-Hilliard equation to the Navier-Stokes equations for fluid flow may uncover behaviour different to that observed for the purely diffusive case.

Experimental work \cite{photod}-\cite{tra} has shown that a variety of mesoscopic structures can be formed when chemical reactions are photo-induced in phase separating polymer mixtures. The effects of two kinds of photo-chemistry have been considered: intermolecular photodimerisations \cite{photod,tra} and intramolecular photoisomerisation \cite{photoi,tra}. Both give rise to a long-range inhibition which prevents phase separation proceeding beyond a certain domain size. In the first case the inhibition is due to the formation of a network of cross-linked polymer molecules whereas in the second case it arises from the differing chemical properties of the two isomers. The similarities in the patterns formed due to phase separation arrest in simple fluids and in reactive polymer blends suggest the latter may be approached by considering first a small-molecule system. 

The paper is organized as follows. In Section~\ref{sec:modelsection} we present a model of a chemically reactive binary fluid which couples the processes of reaction and diffusion to flow. We then outline the linear theory of pattern formation in the absence of hydrodynamic effects. In Section~\ref{sec:LatticeBoltzMethod} we construct a lattice Boltzmann scheme which solves the equations of motion of Section~\ref{sec:modelsection} in the continuum limit. In Sections~\ref{sec:LinearReact} and \ref{sec:QuadReact} results are presented for the evolution of both high and low viscosity systems after a critical quench for a linear and a quadratic reaction mechanism respectively. For the reaction of type $A\leftrightarrow B$, comparison is made with the results of \cite{glotzer}, \cite{christen} and \cite{huo}.

\section{The model}\label{sec:modelsection}
\subsection{Equations of motion}
\label{subsec:eqnMotion}
\hspace{0.5cm}In this section we summarize a model which describes the phase behavior and hydrodynamics of a two-component fluid. Labeling the components $A$ and $B$, we choose a description of the fluid in terms of the following variables: the total density, $\rho=\rho_{A}+\rho_{B}$; the total momentum, $\rho{\bf u}$, and a compositional order-parameter, $\phi=\rho_{A}-\rho_{B}$. 

The composition of the fluid evolves according to a modified version of the Cahn-Hilliard equation which includes the effects of chemical reaction; advection of the order-parameter by the flow-field, {\bf u}, and diffusion in response to gradients in chemical potential:
\begin{equation}
\label{eq:rda}
\partial_{t}\phi({\bf x},t) + \partial_{\alpha}\phi u_{\alpha} =  M_{0}\partial_{\alpha}^{2}\mu({\bf x},t) + J({\bf x},t).
\end{equation}
Here $M_{0}$ is a mobility constant and $J$, which depends on the reaction rate constants, is the change in $\phi$ per unit time due to chemical reactions. The chemical potential of the system, $\mu$, is given by the functional derivative of the free energy, ${\cal F}$, with respect to $\phi$.

We choose a free energy
\begin{equation}
\label{eq:freeEnergy}
{\cal F}[\phi](t) = \int d{\bf x}\left(\frac{\varepsilon}{2}\phi^{2}+\frac{\gamma}{4}\phi^{4}+\frac{\kappa}{2}(\nabla\phi)^{2} + T\rho\ln~\rho \right).
\end{equation}
$\gamma$ is taken to be greater than zero for stability and the sign of $\varepsilon$ determines whether the polynomial contribution to the free-energy density has one or two minima, and hence whether the fluid is above ($\varepsilon>0$) or below ($\varepsilon<0$) its critical temperature. For $\varepsilon<0$ the mixture will separate into two bulk components separated by a narrow, but smooth, interface. The gradient-squared term in $\phi$ associates an energy cost with variations in composition and the parameter $\kappa$ is related to the surface tension and governs the width of the interface between the two phases. The parameter $T$ appears in the isotropic part of the pressure tensor and is related to the degree of incompressibility of the fluid \cite{kendon}. A suitable choice is $T=1/3$.

We consider two types of reactive source term, $J$. A linear source
\begin{equation}
\label{eq:linearSource}
 J \left[ \phi \right]  =  \rho ( \Gamma_{2} - \Gamma_{1} ) - \phi (\Gamma_{1} + \Gamma_{2}),
\end{equation}
corresponding to the reversible chemical reaction $A \leftrightarrow B$.  And a quadratic source
\begin{equation}
\label{eq:quadSource}
 J \left[ \phi \right]  = \frac{1}{2}( \Gamma_{1} + \Gamma_{2} )( \phi - \rho ) \left( \phi - \frac{(\Gamma_{2}-\Gamma_{1})\rho}{\Gamma_{1} + \Gamma_{2}} \right),
\end{equation}
corresponding to the reversible chemical reaction $A + B \leftrightarrow 2B$. The constants $\Gamma_{1}$ and $\Gamma_{2}$ are the rates of the forward and backward reactions respectively. We note that, for a spatially homogeneous system, the linear mechanism has a single stable fixed point whereas the quadratic mechanism has a stable fixed point at $\phi = (\Gamma_{2}-\Gamma_{1})\rho/(\Gamma_{1} + \Gamma_{2})$ and an unstable one at $\phi = \rho$. Here we consider only cases where $\Gamma_{1} = \Gamma_{2}= \Gamma$.

The velocity field obeys a Navier-Stokes equation,
\begin{equation}
\label{eq:ns}
\partial_{t} \rho u_{\alpha} + \partial_{\beta} \rho u_{\alpha} u_{\beta} = -\partial_{\beta} P_{\alpha\beta} + \nu\partial_{\beta} S_{\alpha\beta},
\end{equation}
where $P_{\alpha\beta}$ is the pressure tensor, $\nu$ is the viscosity and $S_{\alpha\beta}$ is the viscous stress tensor. The pressure tensor is derived from the free-energy:
\begin{equation}
\label{eq:pressureRelation}
\partial_{\beta} P_{\alpha\beta} = \rho \partial_{\alpha}\left( \frac{\delta {\cal F}}{\delta \rho} \right) + \phi \partial_{\alpha} \left(  \frac{\delta {\cal F}}{\delta \phi} \right).
\end{equation}
This provides a further coupling between the evolution of $\phi$ and ${\bf u}$ in addition to the advection term in (\ref{eq:rda}). From (\ref{eq:pressureRelation}) and (\ref{eq:freeEnergy}) it follows that
\begin{equation}
P_{\alpha\beta} = \left[T \rho + \phi\partial_{\phi}f_{0} - f_{0} - \kappa \phi \nabla^{2}\phi - \frac{\kappa}{2} (\nabla \phi)^{2} \right] \delta_{\alpha\beta} +\kappa \partial_{\alpha}{\phi} \partial_{\beta}{\phi},
\end{equation} 
where $f_{0}=\varepsilon\phi^{2}/2+\gamma\phi^{4}/4$ denotes the polynomial contribution to the free-energy density.

The total mass density of the fluid $\rho$ is also conserved and obeys
\begin{equation}
\label{eq:consMass}
\partial_{t} \rho + \partial_{\beta} \rho u_{\beta} = 0.
\end{equation}

\subsection{Stability and steady-states without hydrodynamics}\label{basicTheory}

Linear stability analysis of the reaction-diffusion equation (\ref{eq:rda}) with ${\bf u = 0}$ and source term (\ref{eq:linearSource}) shows that only those modes, $\phi({\bf k})$, with $k_{l}<|{\bf k}|<k_{u}$ are unstable, where $k_{u}$ and $k_{l}$ depend on the parameters $\varepsilon$, $\gamma$ and $M_{0}$ in equation (\ref{eq:rda}) \cite{glotzer,christen}. This is in contrast to spinodal decomposition (the case $J=0$) where only short-wavelength modes are stable. The damping of long-wavelength modes in the reactive case prevents continued growth of domains. Instead, phase separation is halted at some length-scale set by the reaction. In addition, there is a threshold value of $\Gamma$,
\begin{equation}
\Gamma_{th} = M_{0}\varepsilon^{2}/8 \kappa,
\end{equation}
above which the reaction is strong enough to completely inhibit phase separation by rendering all linear modes stable.

We also note, following \cite{christen} and \cite{glotzer2}, that there is an equivalence between this behaviour and phase ordering in a system with competing short and long-range interactions. The equivalence can be demonstrated by incorporating the reactive source into an effective free-energy for the system where it appears as a non-local term. Formally, we rewrite (1) as
\begin{equation}
\label{eq:effFreeE}
\partial_{t}\phi({\bf x},t)  + \partial_{\alpha}\phi u_{\alpha} = M_{0}\partial_{\alpha}^{2} \left[ \frac{\delta}{\delta \phi}\left( {\cal F}  - {\cal H} \right) \right],
\end{equation}
where
\begin{equation}
{\cal H} = \frac{\Gamma}{M_{0}}\int d{\bf x}d{\bf y}\phi({\bf x},t)G({\bf x},{\bf y})\phi({\bf y},t),
\end{equation}
and $G({\bf x},{\bf y})$ is the Green's function of the Laplace operator. The reaction is then seen to act as an effective long-range, intra-species repulsion, with strength governed by $\Gamma$, in contrast to the short-range attraction of like-molecules which drives phase separation.

For the case of equal forward and backward rates, the linearised behaviour of the quadratic source (\ref{eq:quadSource}) with reaction rate $\Gamma$ is the same as for the linear source with reaction rate $\Gamma\rho/2$. This can be seen from the linearisation of (\ref{eq:rda}) with ${\bf u=0}$ and source term (\ref{eq:quadSource}). Hence, at early times, the linear and quadratic cases segregate in the same way. However, after the formation of separate A and B-rich regions, non-linear contributions to the source term become important and this is expected to lead to growth of the A-rich phases at the expense of the B-rich ones. This follows from the asymmetry of $J$ between the two phases: in the A-rich phase the production of B is limited by the amount of B already present. Hence it is limited by the {\it  minority} component of the fluid in these regions. In the B-rich regions the production of A is limited by the amount of the majority phase present. Therefore production of A in the B-rich phase is the more rapid process.

\section{A Lattice Boltzmann Scheme}\label{sec:LatticeBoltzMethod}
The lattice Boltzmann method is a well-established numerical technique for hydrodynamic problems \cite{succi}. Initially it was a kinetic-theory based method for the simulation of isothermal ideal flows which was introduced to circumvent some of the problems which rendered its predecessor, lattice-gas cellular automata, impractical. However, it has since been modified and applied to a variety of problems in the simulation of complex fluids. Examples include binary fluids \cite{shan1,swift,giri}, liquid-gas systems \cite{shan2,swift,luo}, liquid-crystals \cite{denn} and colloidal suspensions \cite{ladd}. A lattice Boltzmann scheme for the simulation of two or more species undergoing reaction and diffusion in a moving, viscous solvent was formulated by Dawson {\it et al} \cite{dawson}. In comparision, our model incorporates the thermodynamics of the multi-component fluid via the Cahn-Hilliard equation.

To simulate the binary fluid model described in Section~\ref{sec:modelsection} we utilise the free-energy lattice Boltzmann method of Swift \emph{et al} \cite{swift}. To this end we define two populations of dynamical variables  $\left\{ f_{i}({\bf x}) \right\}_{i=0}^{n}$ and $\left\{ g_{i}(\bf x) \right\}_{i=0}^{n}$ on the sites of a simple lattice in three-dimensions. On each site the variables $f_{i}$ and $g_{i}$ correspond to a velocity direction ${\bf e}_{i}$ for $i=0,1,..,n$. The dynamical variables are referred to as distribution functions since their moments over the velocity set define the macroscopic physical quantities:
\begin{equation}
\label{eq:moments}
\rho = \sum_{i=0}^{n} f_{i}({\bf x},t)\mbox{,} \;\;\;\;\; \rho{\bf u}= \sum_{i=0}^{n} {\bf e}_{i} f_{i}({\bf x},t)\mbox{,} \;\;\;\;\; \phi = \sum_{i=0}^{n} g_{i}({\bf x},t).
\end{equation}

\begin{figure}
\begin{center}
\includegraphics[angle=270,width=60mm]{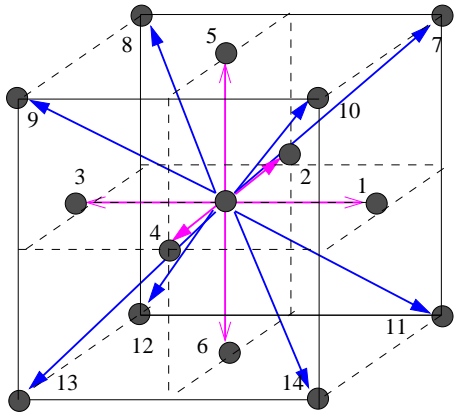}  
\end{center}
\caption{ (Color online) The fifteen-velocity face-centered cubic lattice. The dots represent lattice sites and arrows the velocity vectors used in the lattice Boltzmann scheme. The lattice spacing, $\delta x$, and time-step, $\delta t$, are set to unity. }
\label{fig:lattice}
\end{figure}

The distribution functions on each site are updated in discrete time with a time-step $\delta t$. The velocities are chosen so that ${\bf e}_{0}=0$ and, for all $i\neq 0$, ${\bf e}_{i}\delta t$ lies between two lattice sites. The choice of lattice and velocity set are subject to certain restrictions \cite{succi}. For this work we used a face-centred cubic lattice in three dimensions with the set of fifteen velocities, $i=0,1,..,14$, illustrated in Figure \ref{fig:lattice}.

The distribution functions $f_{i}$ and $g_{i}$ evolve according to
\begin{gather}
\label{eq:evolf} f_{i}({\bf x}+\delta t {\bf e}_{i};t+\delta t) = f_{i}({\bf x};t) + \Delta_{i}^{\tau_{f}}[f], \\
\label{eq:evolg} g_{i}({\bf x}+\delta t {\bf e}_{i};t+\delta t) = g_{i}({\bf x};t) + \Delta_{i}^{\tau_{g}}[g] + F_{i},
\end{gather}
where
\begin{equation}
\Delta_{i}^{\tau}[f] = - \frac{\delta t}{\tau} \left( f_{i}({\bf x};t) - f_{i}^{eq}({\bf x};t) \right),
\end{equation}
and the relaxation times $\tau_{f}$ and $\tau_{g}$ are free parameters. Equations (\ref{eq:evolf}) and (\ref{eq:evolg}) are both lattice equivalents of the Bhatnager-Gross-Krook (BGK), or single relaxation time, approximation to the full continuum Boltzmann equation \cite{succi}.

We now need to specify the local equilibria functions $f_{i}^{eq}$ and $g_{i}^{eq}$ and the forcing term $F_{i}$. $f^{eq}_{i}$ and $g^{eq}_{i}$ are taken to be series expansions in the velocity
\begin{gather}
\label{eq:fequil} f_{i}^{eq} = A_{i} + B_{i}e_{i\alpha}u_{\alpha} + C_{i}u^{2} + D_{i}e_{i\alpha}e_{i\beta}u_{\alpha}u_{\beta} +  E_{i}^{\alpha\beta}e_{i\alpha}e_{i\beta}, \\
\label{eq:gequil} g_{i}^{eq} = H_{i} + J_{i}e_{i\alpha}u_{\alpha} + K_{i}u^{2} + Q_{i}e_{i\alpha}e_{i\beta}u_{\alpha}u_{\beta}.
\end{gather}
The coefficients in (\ref{eq:fequil}) and (\ref{eq:gequil}) are chosen so the moments of the equilibrium distributions satisfy
 \begin{gather}
       \sum_{i=0}^{n} f^{eq}_{i} = \rho, ~ ~ ~  \sum_{i=0}^{n} f^{eq}_{i}e_{i\alpha} = \rho u_{\alpha} \label{eq:fmoments}, \\     
       \sum_{i=0}^{n} f^{eq}_{i}e_{i\alpha}e_{i\beta}  = P_{\alpha \beta} + \rho u_{\alpha} u{_\beta} \label{eq:secondfmoment},\\
       \sum_{i=0}^{n} g^{eq}_{i} = \phi, ~ ~ ~  \sum_{i=0}^{n} g^{eq}_{i}e_{i\alpha} = \phi u_{\alpha} \label{eq:gmoments}, \\
       \sum_{i=0}^{n} g^{eq}_{i}e_{i\alpha}e_{i\beta} =D\mu\delta_{\alpha \beta}+\phi u_{\alpha}
                             u{_\beta} \label{eq:secondgmoments}.
 \end{gather}
In addition, the lattice forcing term is chosen to obey
\begin{gather}
\label{eq:Fmoments}
\sum_{i=0}^{n} F_{i} = \delta t J[\phi], \;\;\; \sum_{i=0}^{n} e_{i\alpha} F_{i} = 0.
\end{gather}

One possible choice of the coefficients, such that constraints (\ref{eq:fmoments})-(\ref{eq:Fmoments}) hold, is given by
\begin{gather}
A_{1-14} = \frac{1}{30} P_{\alpha\alpha},\;\;\; A_{0} = \rho - 14A_{1}, \\
B_{7-14} = \rho/24,\;\;\; B_{1-6} = 8B_{7}, \\
C_{7-14} = -\rho/24,\;\;\; C_{0} = 16C_{7},\;\;\; C_{1-6} = 2C_{2}, \\
D_{7-14} = \rho/16,\;\;\; D_{1} = 8D_{2}, \\
E_{7-14}^{\alpha\beta} = \frac{1}{16}\left[ P_{\alpha\beta} - \frac{1}{3}\delta_{\alpha\beta} P_{\gamma\gamma} \right], \\
E_{1-6}^{\alpha\beta} = 8E_{7}^{\alpha\beta}, \\
H_{1-14} = \frac{1}{10}D\mu,\;\;\; H_{0} = \phi - 14H_{1}, \\
J_{7-14} = \phi/24,\;\;\; J_{1-6} = 8J_{7}, \\
K_{7-14} = -\phi/24,\;\;\; K_{0} = 16K_{7}, \;\;\; K_{1-6} = 2K_{7}, \\
Q_{7-14} = \phi/16,\;\;\; Q_{1-6} = 8Q_{7} \\
F_{0} = \delta t\frac{1}{2}J,\;\;\; F_{1-6} = \delta t\frac{1}{24}J,\;\;\; F_{7-14} = \delta t\frac{1}{32}J.
\end{gather}

The constraints (\ref{eq:fmoments})-(\ref{eq:Fmoments}) ensure that, on length and time-scales large compared to the lattice-spacing and time-step, the evolution of the moments (\ref{eq:moments}) satisfies the partial differential equations set out in Section~\ref{subsec:eqnMotion}. To check that this is indeed the case the task of reducing the description of the dynamics in terms the distribution functions to one in terms of their moments must be addressed. The reduction can be performed by a Chapman-Enzkog expansion of equations (\ref{eq:evolf}) and (\ref{eq:evolg}). Since the details of this are essentially no different to those found in \cite{swift} we present only the result here.

The zeroth moment of the $f_{i}$ satisfies equation (\ref{eq:consMass}) for conservation of mass. The first moment of the $f_{i}$ satisfies the Navier-Stokes equation (\ref{eq:ns}) with $\nu=\frac{1}{3}\left( \tau_{f} - 1/2 \right)\delta t$ and
\begin{equation}
S_{\alpha\beta} = \rho (\partial_{\alpha} u_{\beta} + \partial_{\beta} u_{\alpha}) + {\cal E}(\rho,{\bf u}),
\end{equation}
where the ${\cal E}$ denotes unwanted error terms \cite{swift}.

The first moment of the $g_{i}$ satisfies the reaction-diffusion-advection equation
\begin{eqnarray}
\partial_{t} \phi + \partial_{\alpha} \phi u_{\alpha}& = & \omega_{g} \left[D \partial_{\alpha}^{2} \mu - \partial_{\beta}\left( \frac{\phi}{\rho}\partial_{\alpha}P_{\alpha\beta} \right) \right] + \omega_{g}\partial_{\alpha} \left( u_{\alpha} J \right) + J[\phi],
\label{eq:rda2}
\end{eqnarray}
where $\omega_{g}= \delta t(\tau_{g}-1/2)$ and $D$ is defined in (\ref{eq:secondgmoments}). Equation (\ref{eq:rda2}) corresponds to equation (\ref{eq:rda}) with mobility $M_{0}=\omega_{g}D$ but with two extra terms. The term in gradients of the components of the pressure tensor is present in some other free-energy lattice Boltzmann methods and has been shown numerically to be small in comparison to the desired terms \cite{swift}. The term in gradients of the reactive source can be seen, to first order in $\phi$, to be a correction to the advecting velocity of order $\omega_{g}\Gamma$. For our choices of parameters, this contribution is small in the low viscosity regime where the advecting flow field is important.

This completes the specification of our lattice Boltzmann method. Although the model is inherently three-dimensional, we consider only its restriction to two dimensions. As an initial condition we choose the total density of the fluid $\rho=1$ at each lattice lattice. The near-incompressibility of the fluid ensures that this value remains approximately the same at later times. To imitate the conditions following a rapid cooling of a fluid from above to below its critical temperature we initialise $\phi({\bf x};0)=\delta\phi({\bf x})$ where $\delta\phi$ is random noise with $ | \delta\phi ({\bf x}) | \leq 0.01 \;\; \forall {\bf x}$. The parameters $\varepsilon$ and $\gamma$ in the free-energy are chosen so that in the unreactive case the fluid phase separates into regions where $\phi=\pm 1.0$. We choose $\kappa=0.01$ to ensure a narrow interface and choose $\tau_{g}=1.0$ and $D=2.0$ which fixes the diffusion constant $M_{0}=1.0$. We also choose $\delta t = \delta x = 1.0$ and the system size $L_{x}=L_{y}=256$, throughout. The viscosity of the fluid is controlled by varying the relaxation time $\tau_{f}$. For $\tau_{f}=400$, domains in the unreactive fluid grow as $t^{1/3}$ and the fluid can be taken to be in the diffusive regime where hydrodynamic flows are negligible. To ensure a hydrodynamic growth exponent of $2/3$ we choose $\tau_{f}=5.0$.

\section{Linear Reaction}\label{sec:LinearReact}

\begin{figure}
\centering
\begin{tabular}{ccc}
\epsfig{file=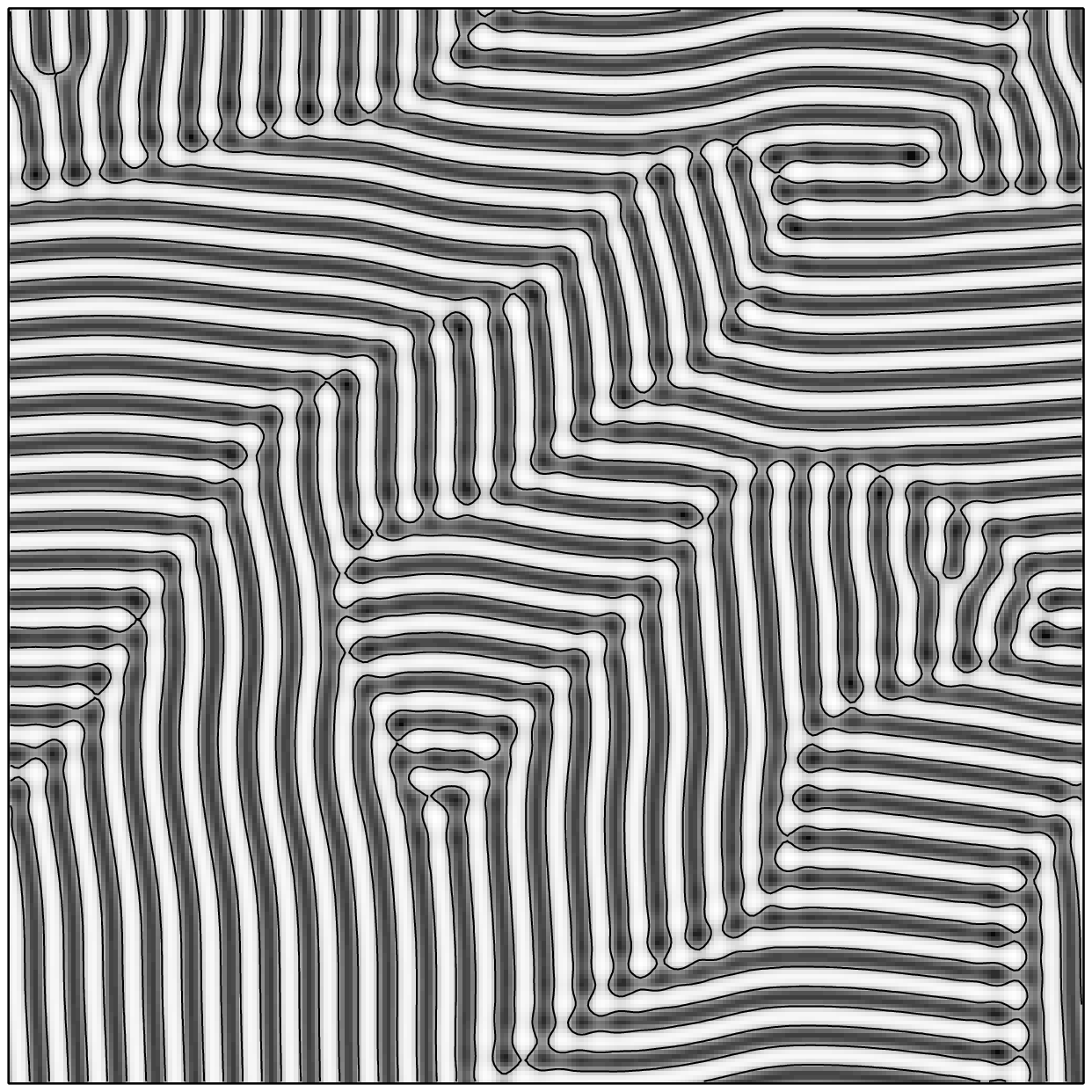,width=40mm} & \epsfig{file=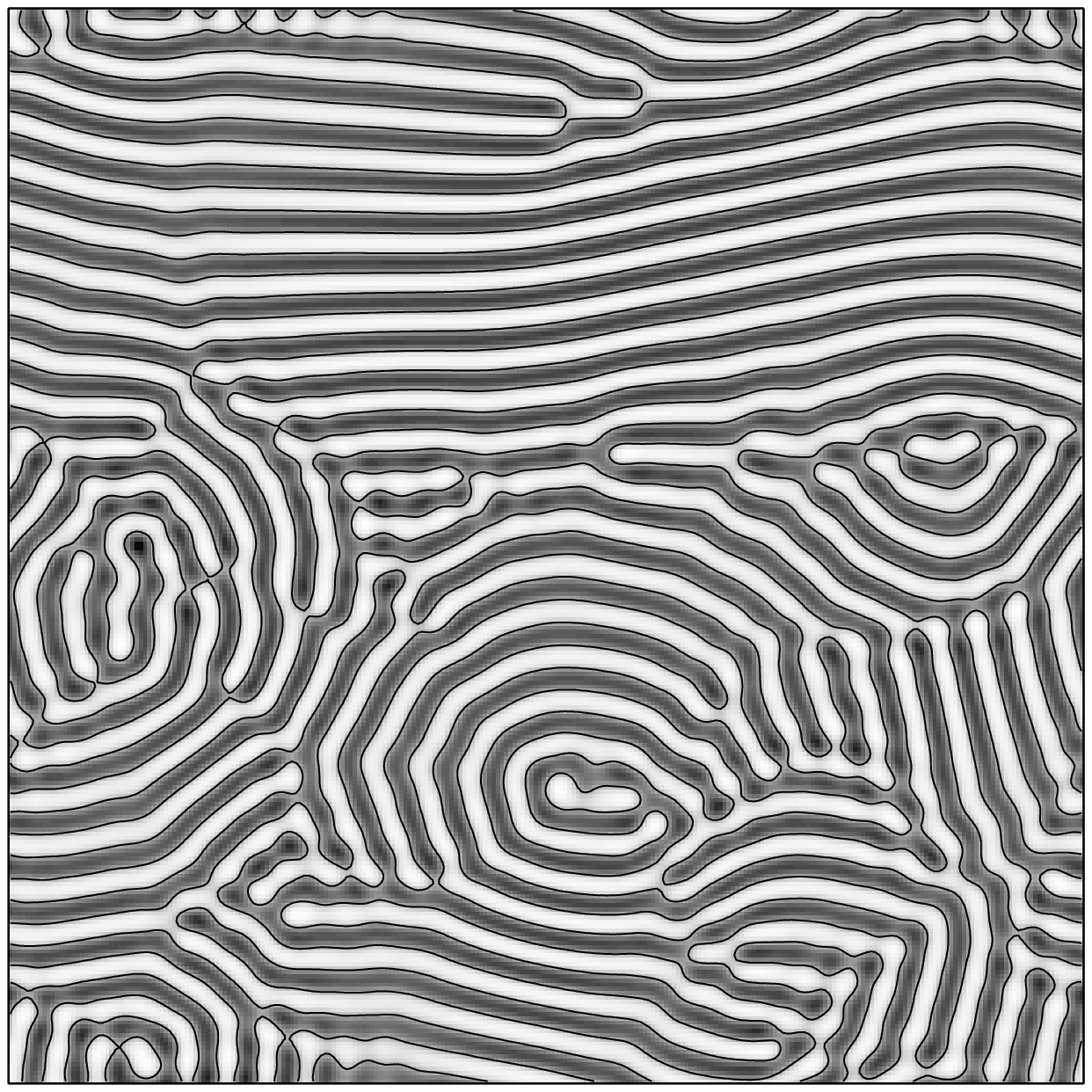,width=40mm} \\
\epsfig{file=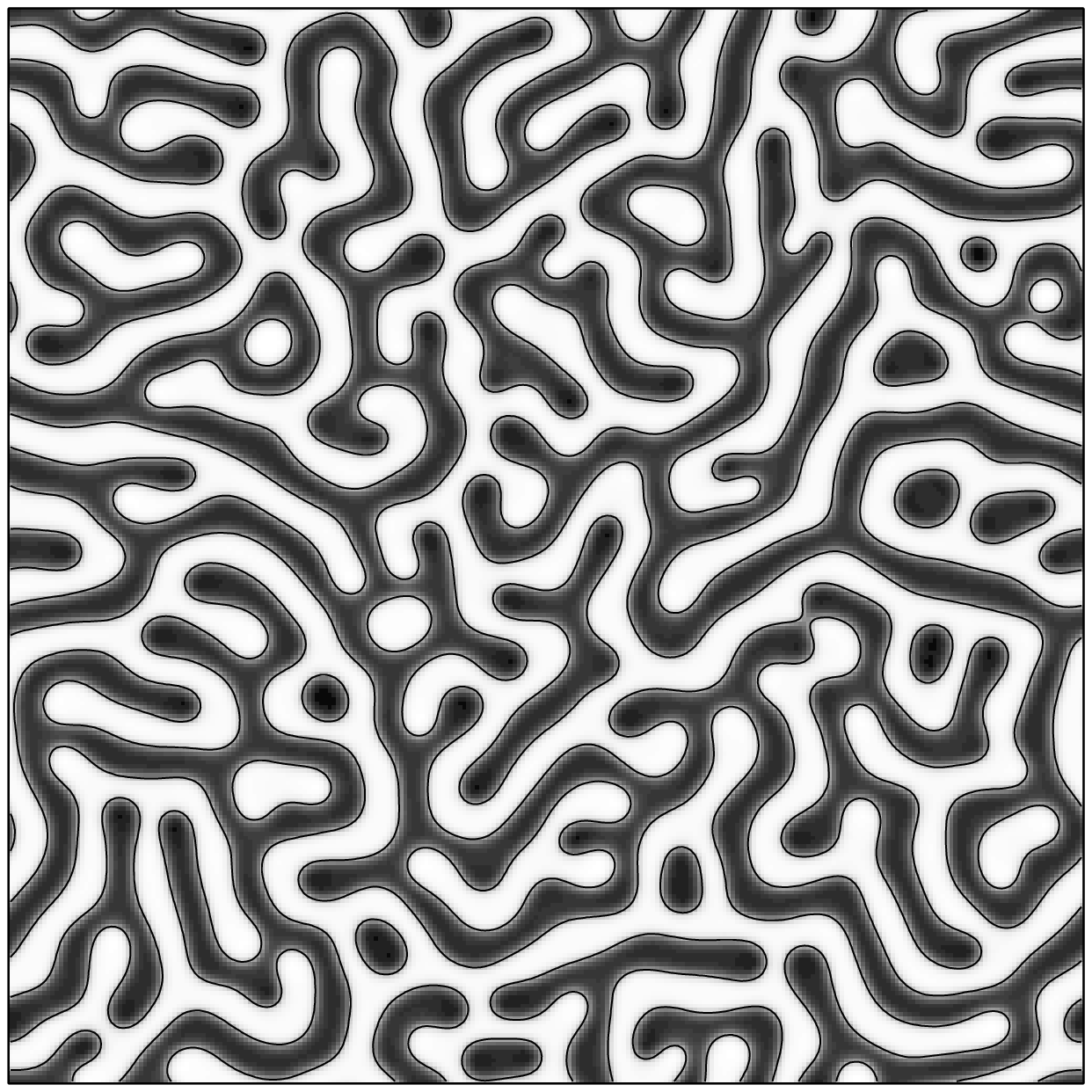,width=40mm} & \epsfig{file=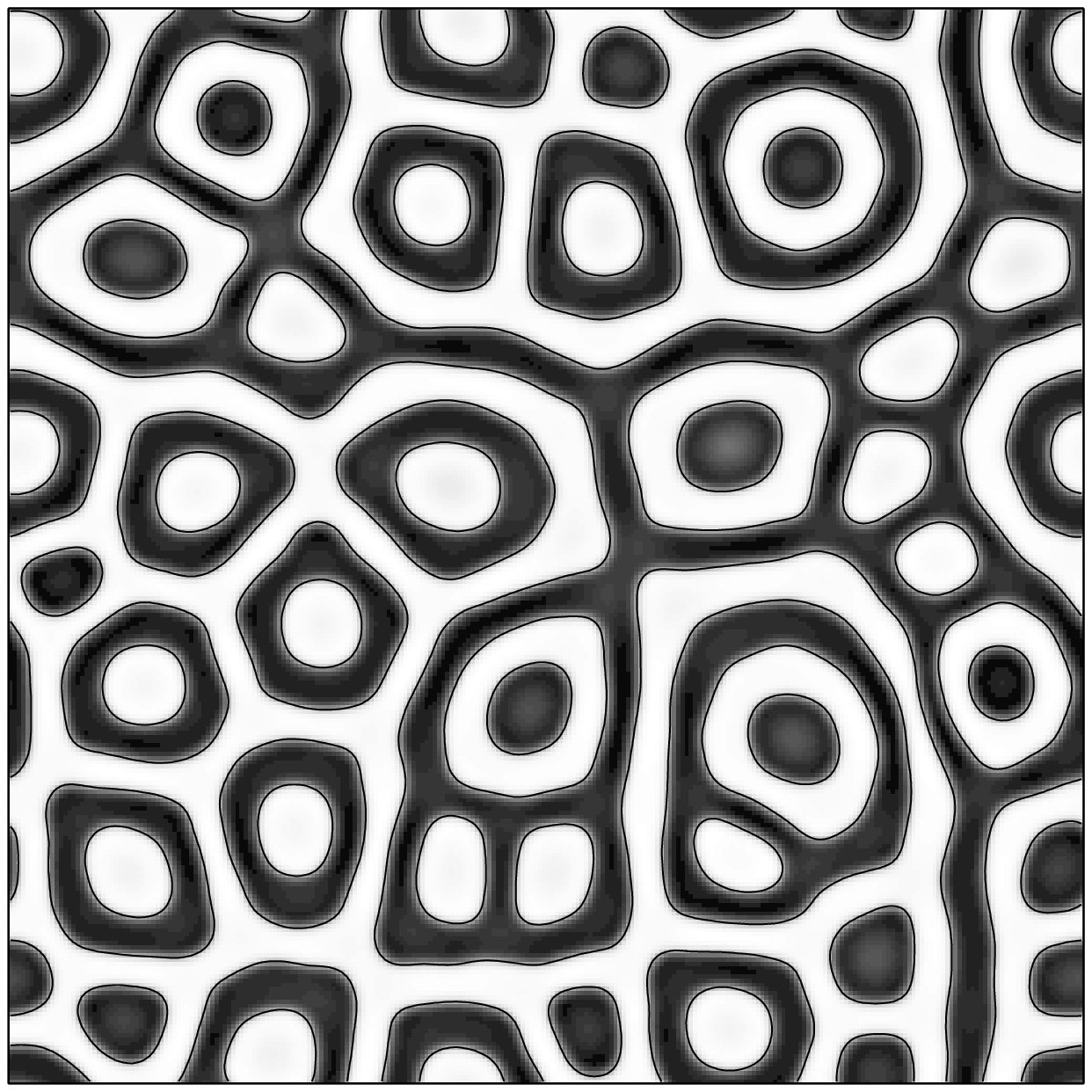,width=40mm}
\end{tabular}
\caption{Steady, or near-steady, states for a linear reaction mechanism in the diffusive ($\tau_{f}=400$, left-hand column) and hydrodynamic ($\tau_{f}=5.0$, right-hand column) regimes for two different values of $\Gamma$: $\Gamma=0.001$, top, and $\Gamma=0.0001$, bottom. $A$-rich regions are shown in white and $B$-rich ones in black.}
\label{fig:LsteadyStates}
\end{figure}

We consider first the case of the linear reaction mechanism (\ref{eq:linearSource}). Figure \ref{fig:LsteadyStates}
 compares the steady-state structures at low and high viscosities for two different values of the rate constant $\Gamma$. 

At high viscosity (Fig.~\ref{fig:LsteadyStates}, left-hand column) hydrodynamic effects are suppressed and domain evolution is due entirely to the interplay between reaction and diffusion. We observe the laminar steady-state morphologies found in \cite{glotzer}, \cite{christen}. The average domain size in the steady-state is dependent on the rate of reaction: decreasing $\Gamma$ weakens the opposition to phase separation posed by the reaction. Hence, for smaller $\Gamma$, the average domain size in the steady-state is larger.

\begin{figure}
\psfrag{L}{$l_{I}$}
\psfrag{t}{$t$}
\centering
\begin{tabular}{ccccc}
\epsfig{file=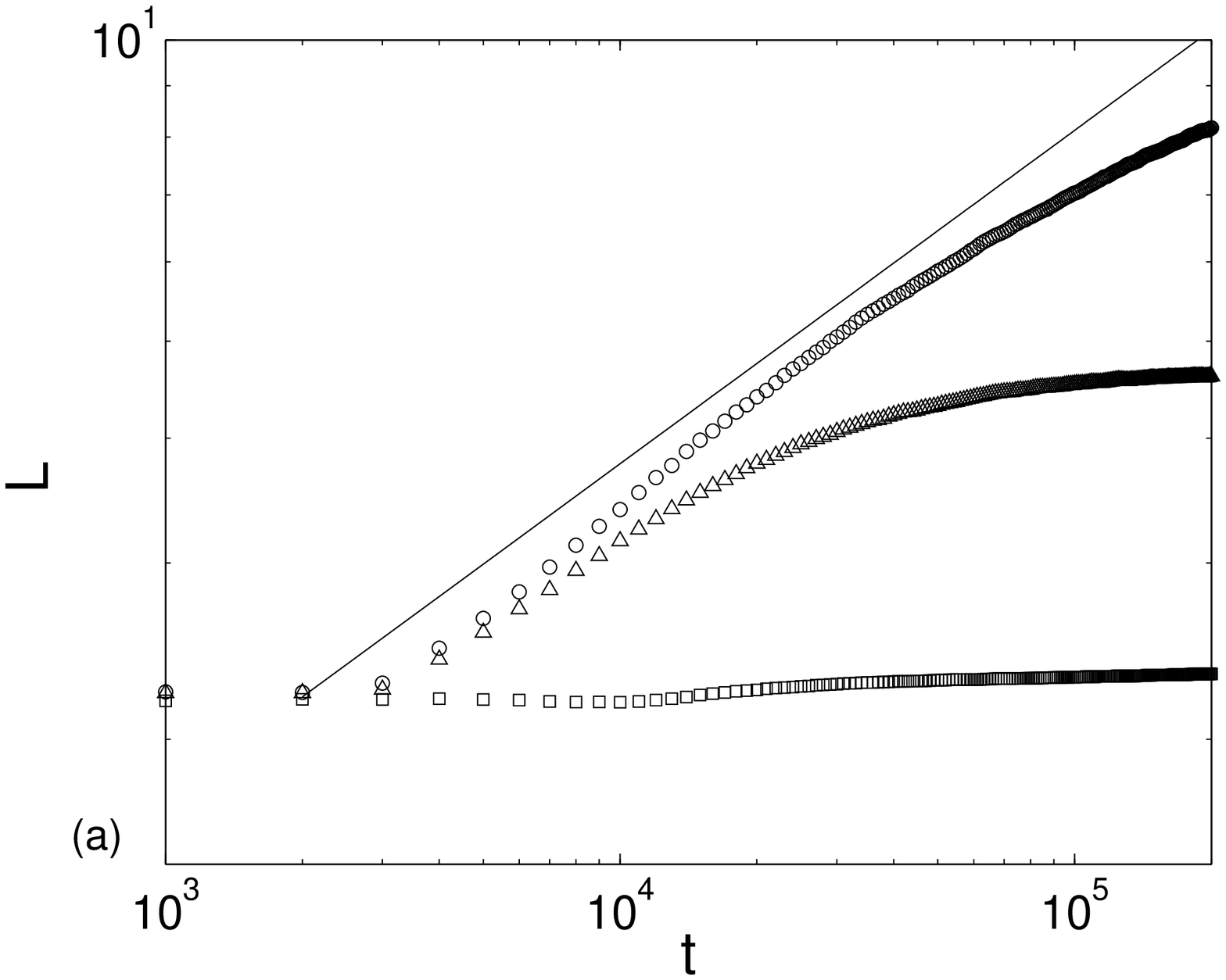,width=55mm} & & & & \epsfig{file=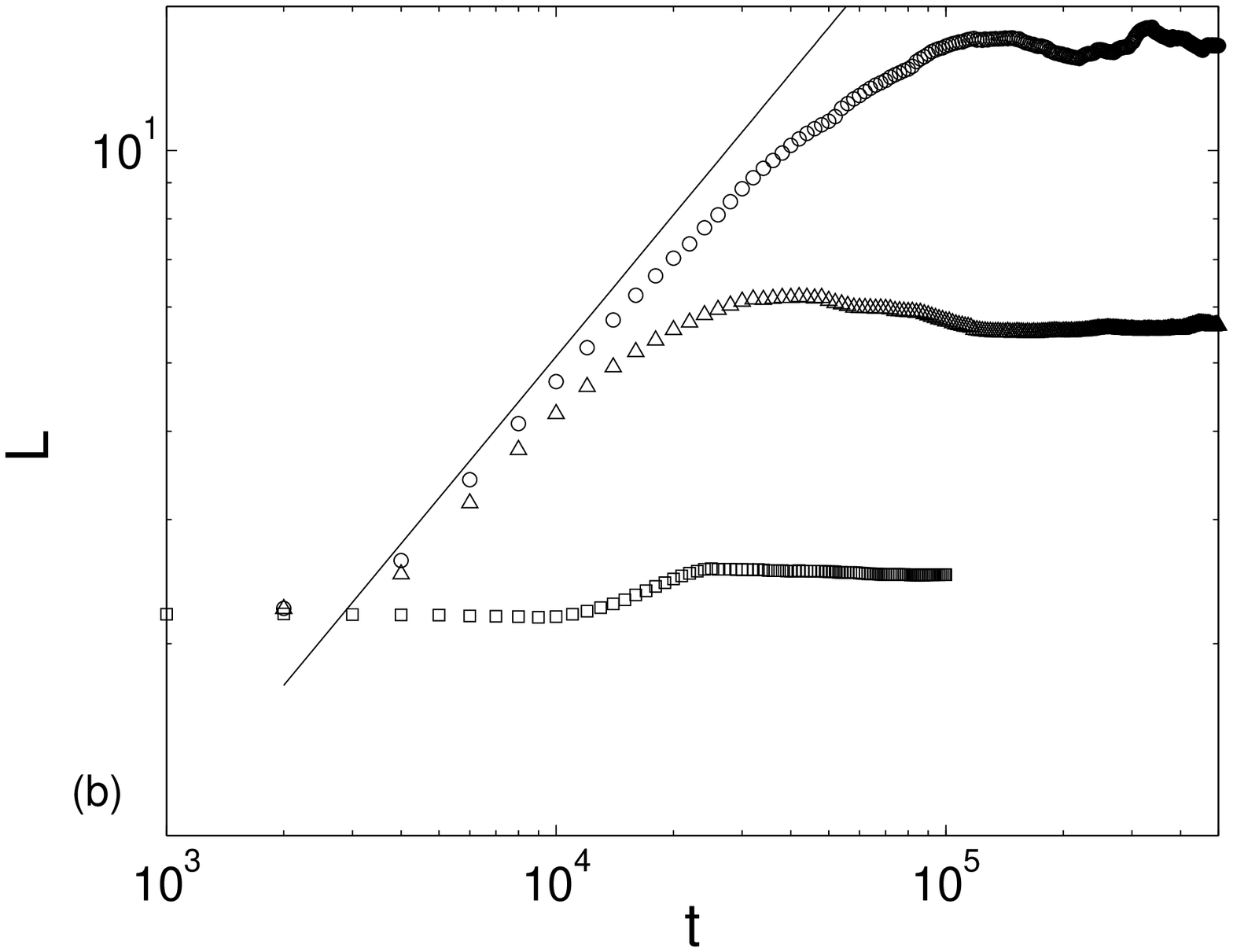,width=55mm}
\end{tabular}
\caption{ Log-log plot of the average domain size, measured by the inverse interfacial length, as a function of time at (a)~high and (b)~low viscosity. The reaction rates are $\Gamma=0.001$ ($\square$); $0.0001$ ($\triangle$) and $10^{-5}$ ($\circ$), respectively. The solid straight line corresponds to $\alpha=1/3$ in (a) and $\alpha=2/3$ in (b).}
\label{fig:lRLengthScale}
\end{figure}

A measure of the characteristic length-scale is given by the inverse interfacial length $l_{I} = L_{x}L_{y}/L_{I}$, where $L_{I}$ is defined as the number of lattice sites ${\bf x}$ such that $\phi({\bf x})\phi({\bf x}^{'})<0$ for at least one ${\bf x}^{'}$ a nearest neighbour of ${\bf x}$. Figure \ref{fig:lRLengthScale}(a)
 shows the time-evolution of $l_{I}$ at high viscosity for different values of $\Gamma$. For low rate constants, $l_{I}$ is close to the spinodal growth-law, $\alpha =1/3$, at early times before asymptoting to a constant value. For a faster reaction, a steady length-scale is quickly selected without prior scaling. These results are in qualitative agreement with those of \cite{christen} and \cite{huo}.

At low viscosities hydrodynamics becomes important and pressure driven flows attempt to eliminate local interfacial curvature. The effect on phase separation in an unreactive fluid is to increase the rate of domain growth. However, the mechanism is such that for a complicated domain structure, conflicting signals will be sent into the bulk phase as to the direction in which flow should be established \cite{siggia}. As a result, a complicated flow-field arises.

For $\Gamma=0.001$ a labyrinthine pattern of A and B-rich domains, similar to that observed at high viscosity, is formed (compare the top-two panels in Fig.~\ref{fig:LsteadyStates}). However, the regions of highest interfacial curvature at high viscosity (Fig.~\ref{fig:LsteadyStates}, top-left), where two mis-aligned domains meet, can no longer be maintained in the presence of flow. Instead the domains curve gently and form concentric arrays of arcs, sometimes enclosing spiral-like structures \cite{huo}. 

For $\Gamma=0.0001$, the additional mixing effects of pressure driven flows lead to a greater degree of coarsening before the energetic barrier presented by the reaction is reached than in the purely diffusive case (compare the bottom-two panels in Fig.~\ref{fig:LsteadyStates}). Hence, the domains at low viscosity are larger and more circular compared to the narrow, elongated domains found at high viscosity. At later times there is a period of domain {\it creation} with smaller domains forming inside existing ones. This is analogous to the `interface-induced secondary quench' described by Tanaka \cite{tan}. Rapid hydrodynamic growth has resulted in a state where the processes of reaction and diffusion have not yet brought about bulk phases with equilibrium values of $\phi$. Consequently inhomogeneities in the bulk may grow, triggering the formation of new domains within an existing one. Figure~\ref{fig:lRLengthScale}(b) shows the evolution of the inverse interfacial length at low viscosity for three different reaction rates. For $\Gamma=10^{-4}$ and $\Gamma=10^{-5}$, the introduction of new interface is clearly seen as a peak in $l_{I}$. At late times a balance is eventually struck between the flow, diffusion and the reactive repulsion. The result is the hierarchy of roughly circular interfaces separated by narrow channels shown in Fig.~\ref{fig:LsteadyStates}(bottom-right).

\section{Quadratic Reaction}\label{sec:QuadReact}
\begin{figure}
\centering
\begin{tabular}{ccc}
\epsfig{file=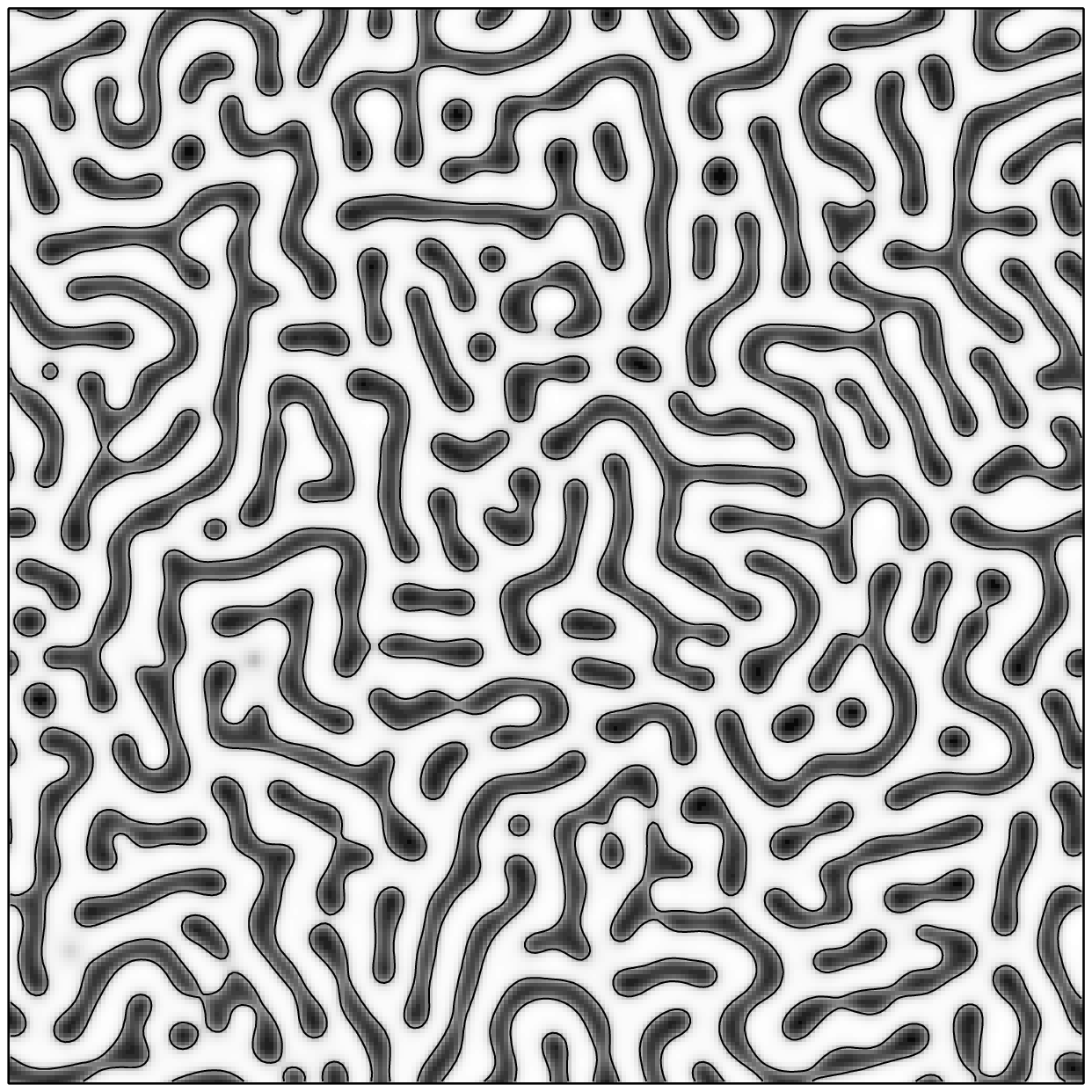,width=40mm} & \epsfig{file=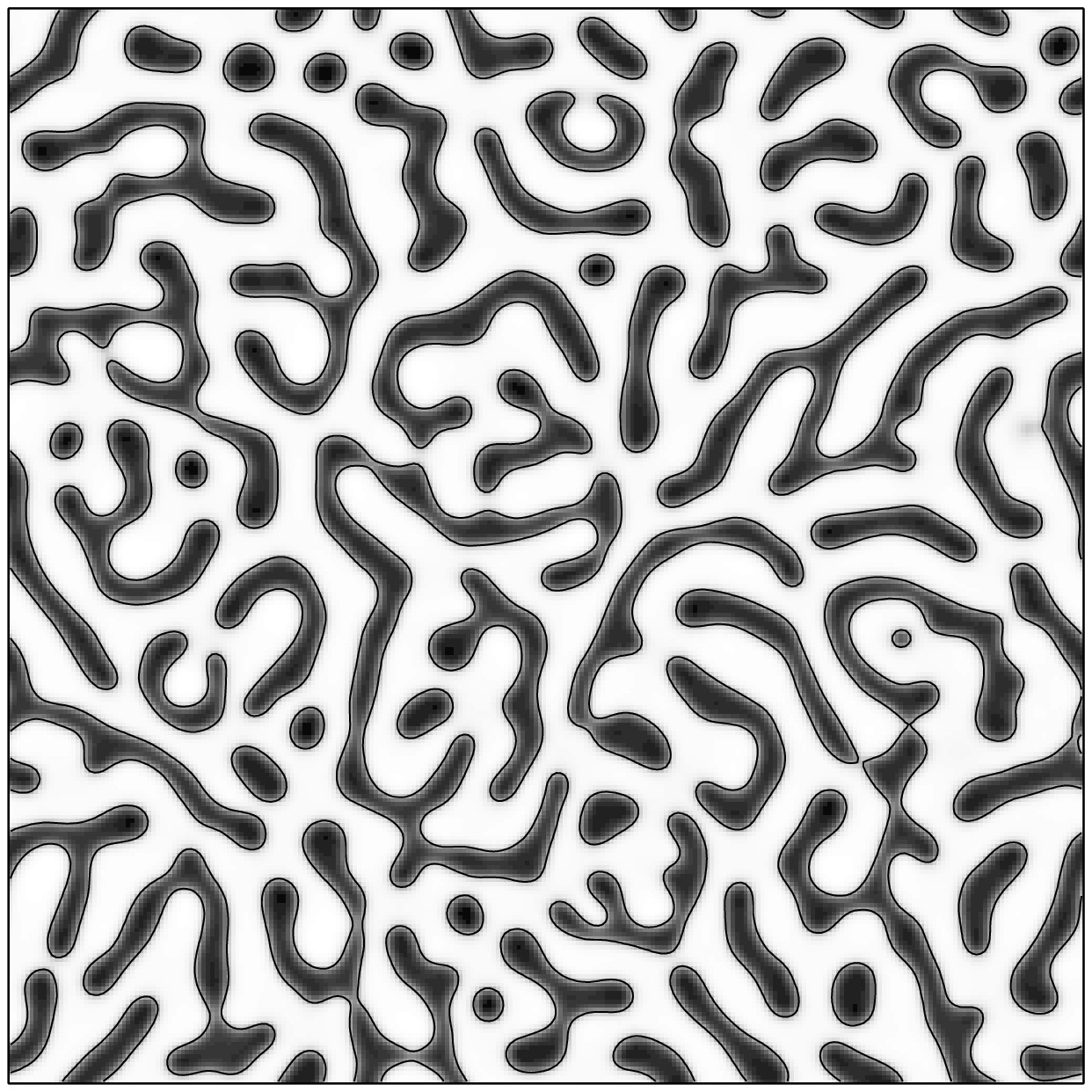,width=40mm} \\
\epsfig{file=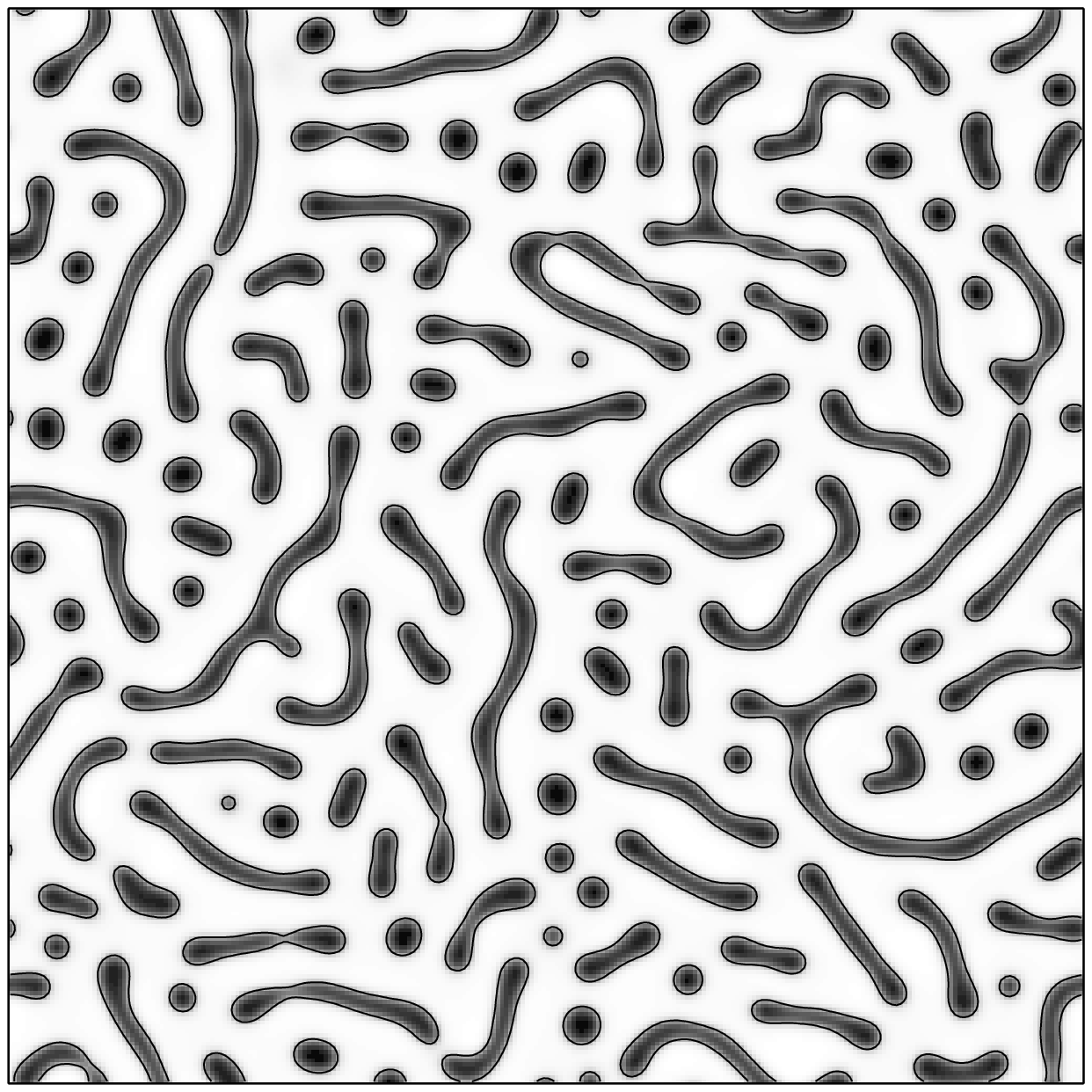,width=40mm} & \epsfig{file=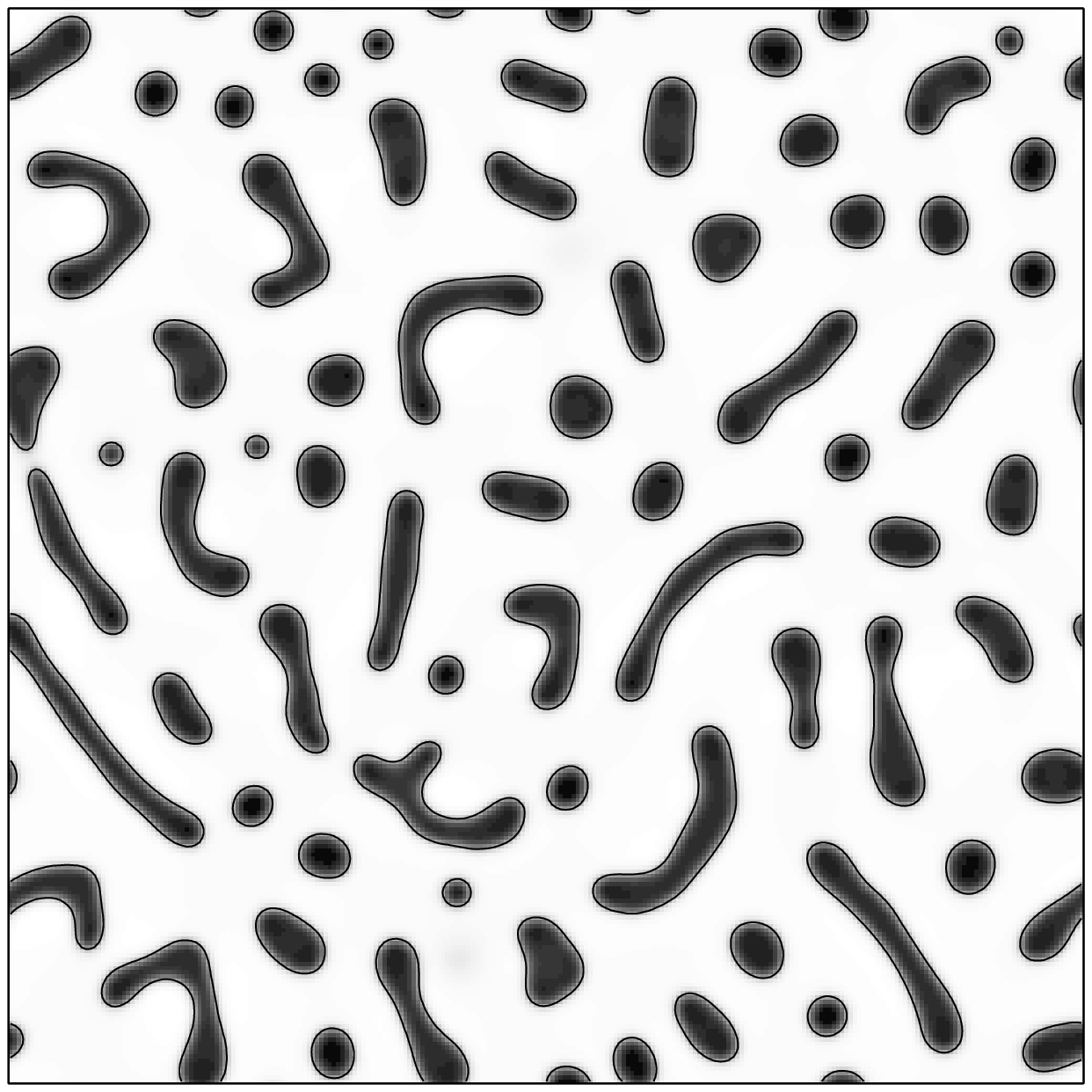,width=40mm} \\
\epsfig{file=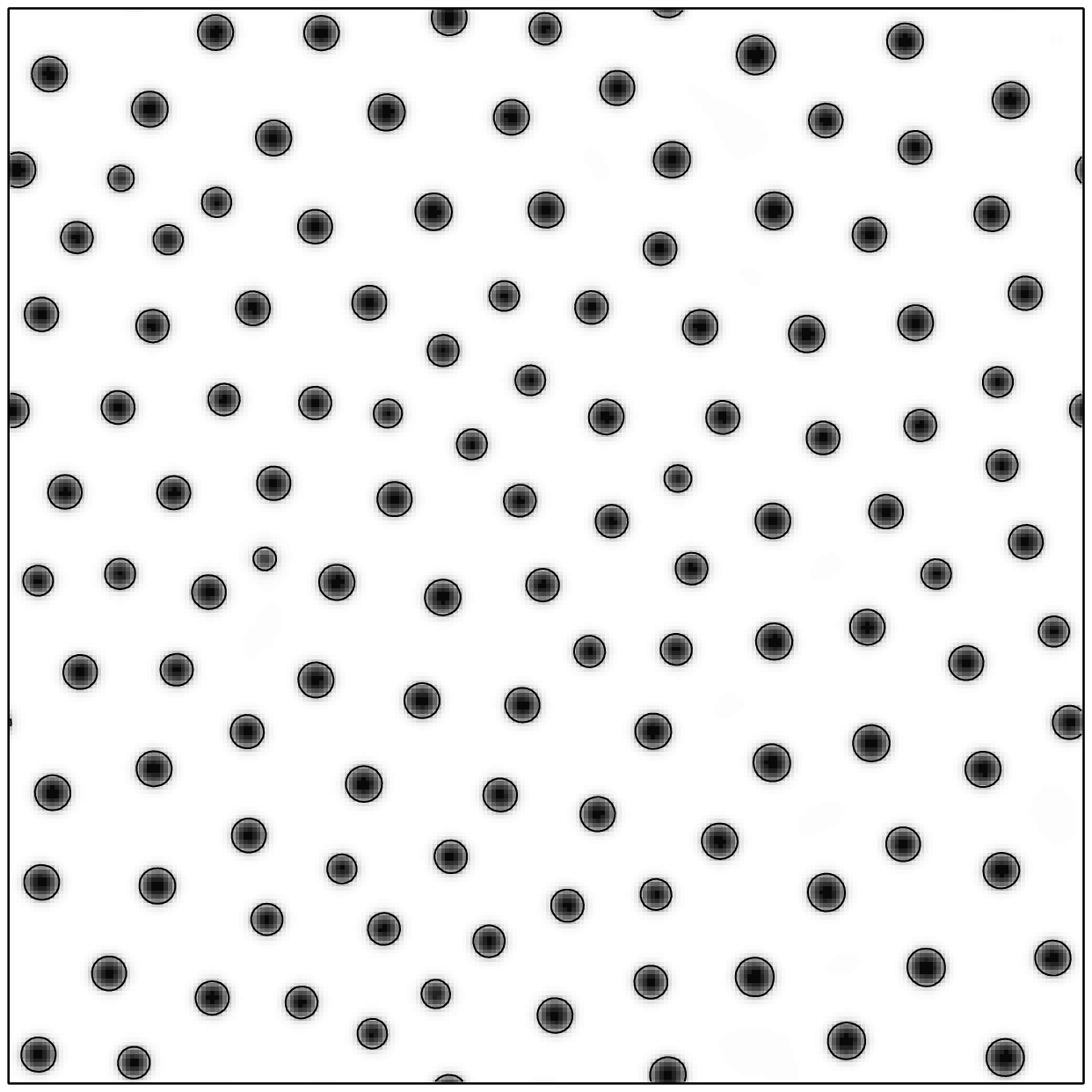,width=40mm} & \epsfig{file=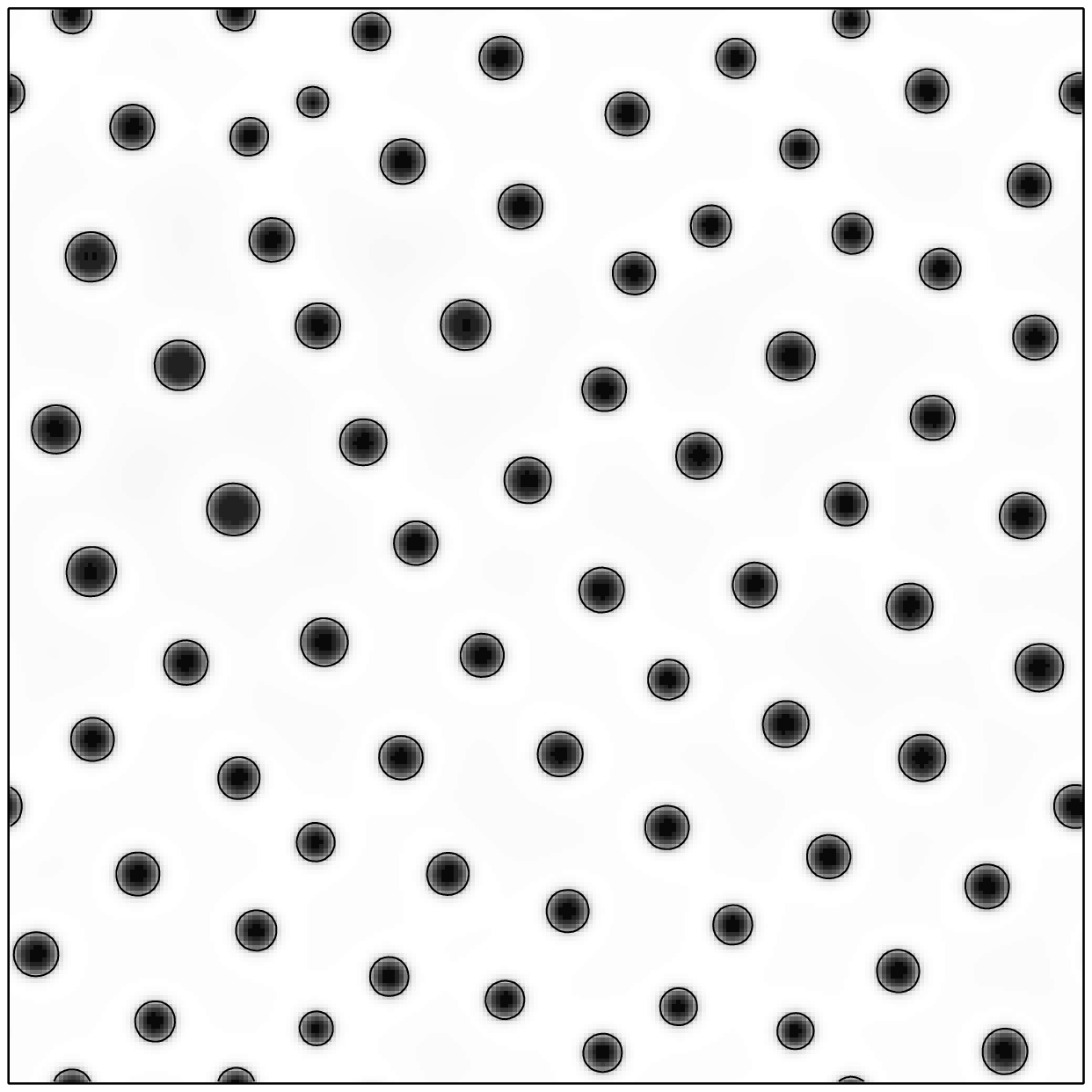,width=40mm}
\end{tabular}
\caption{ Time evolution of the domain structure for a quadratic reaction mechanism with rate $\Gamma=0.0001$ for high viscosity ($\tau_{f}=400$, left-hand column) and low viscosity ($\tau_{f}=5.0$, right-hand column) at times $t=10,000$ (top), $t=20,000$ (middle) and $t=200,000$ (bottom). The $A$-rich regions are shown in white and the $B$-rich regions in black.}
\label{fig:qRslowRate}
\end{figure}

For the case of the quadratic source term (\ref{eq:quadSource}), the steady state is an array of $B$-rich domains in an $A$-rich matrix where, in both phases, the order-parameter does not take a thermodynamic equilibrium value. The shape of the steady-state domains, their density and the evolution to the steady-state depend on the viscosity and the rate of reaction. The left-hand column of Figure~\ref{fig:qRslowRate}
\begin{figure}
\centering
\begin{tabular}{ccc}
\epsfig{file=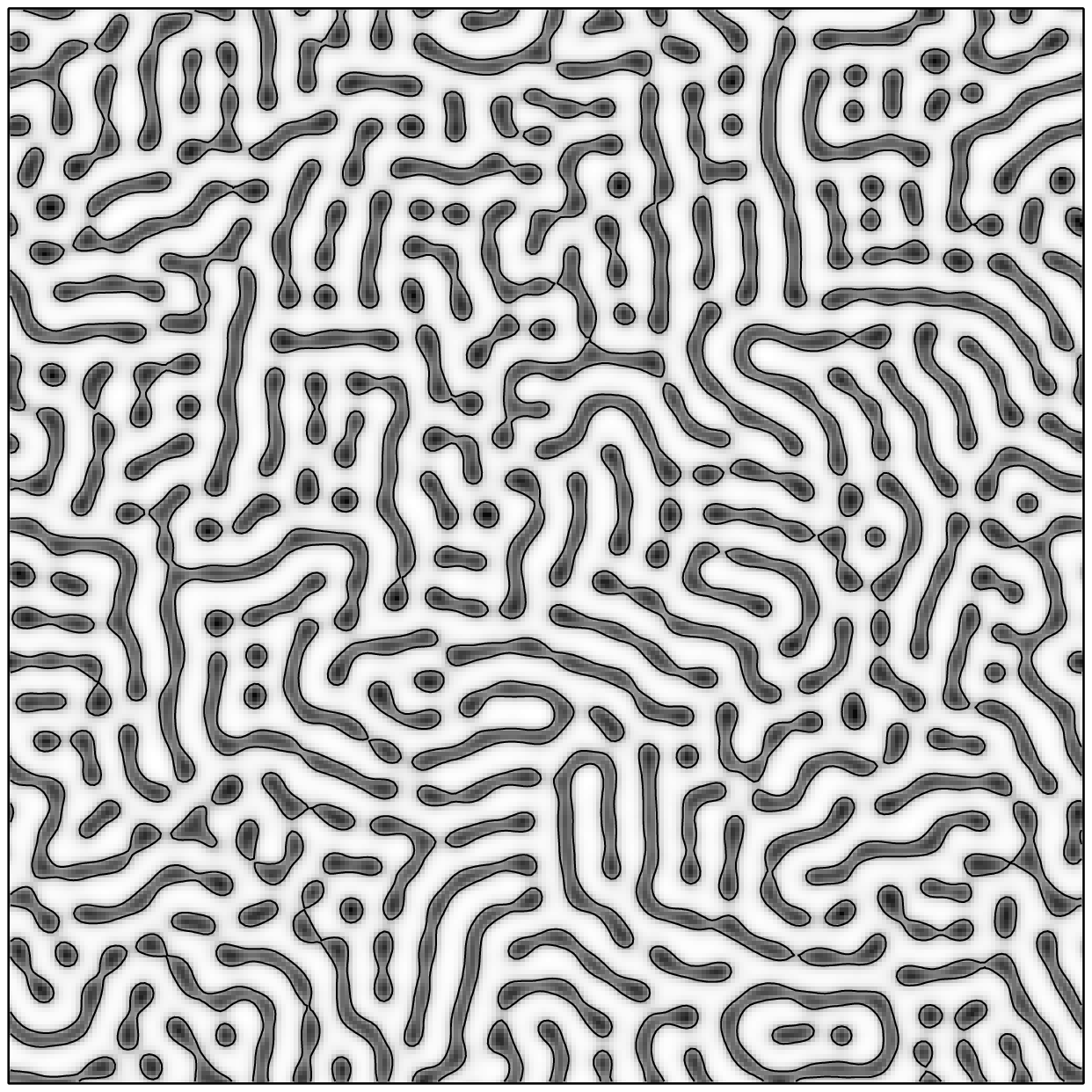,width=40mm} & \epsfig{file=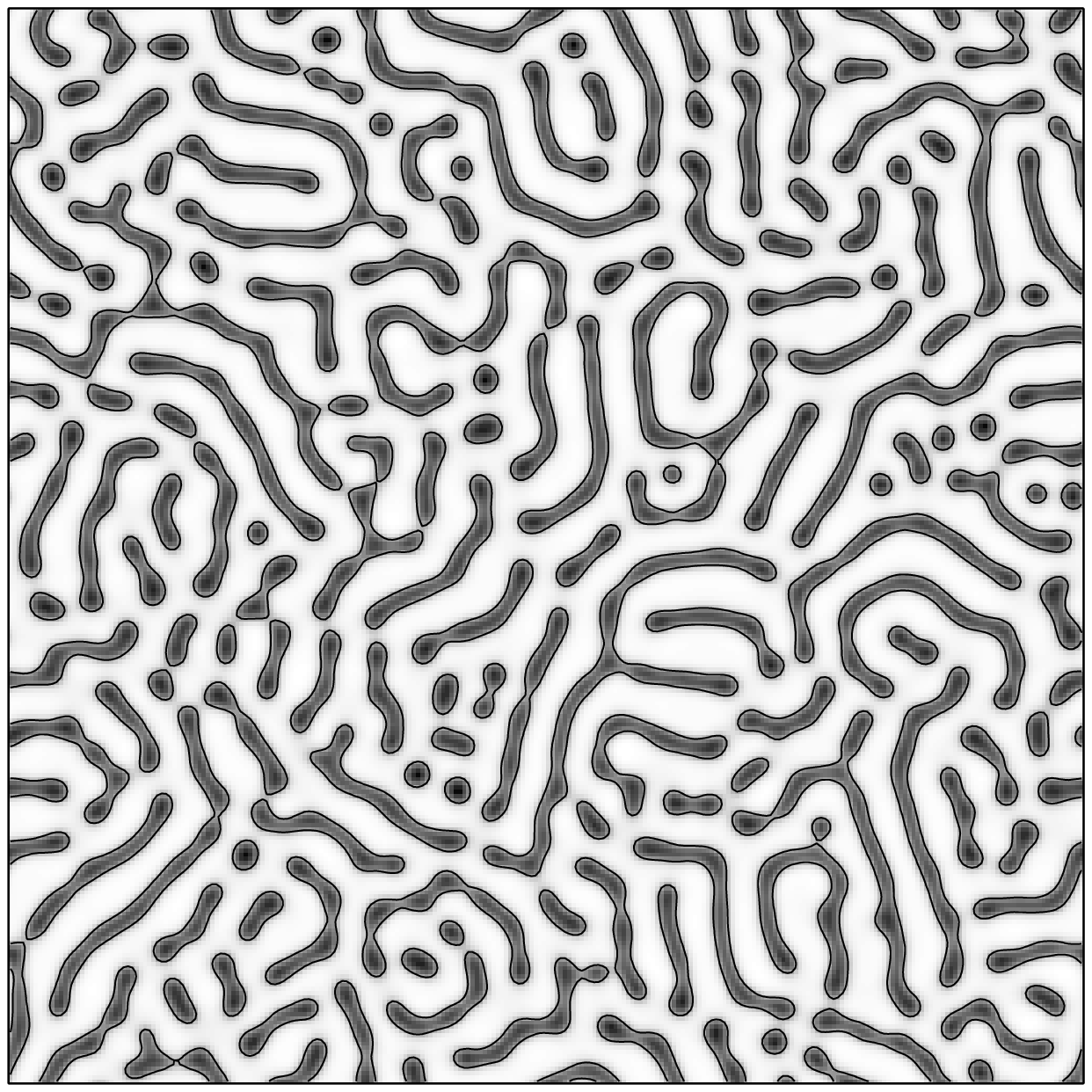,width=40mm} \\
\epsfig{file=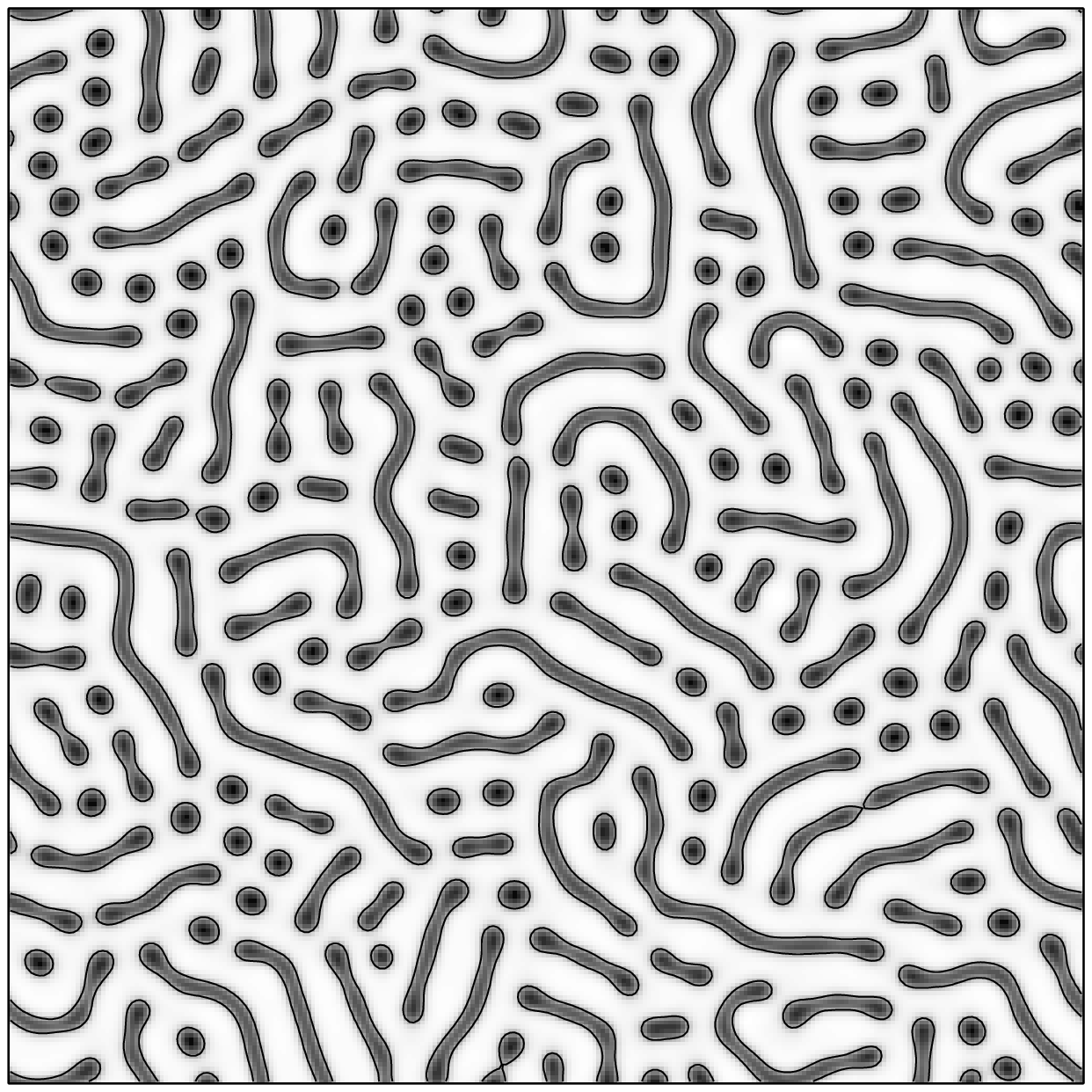,width=40mm} & \epsfig{file=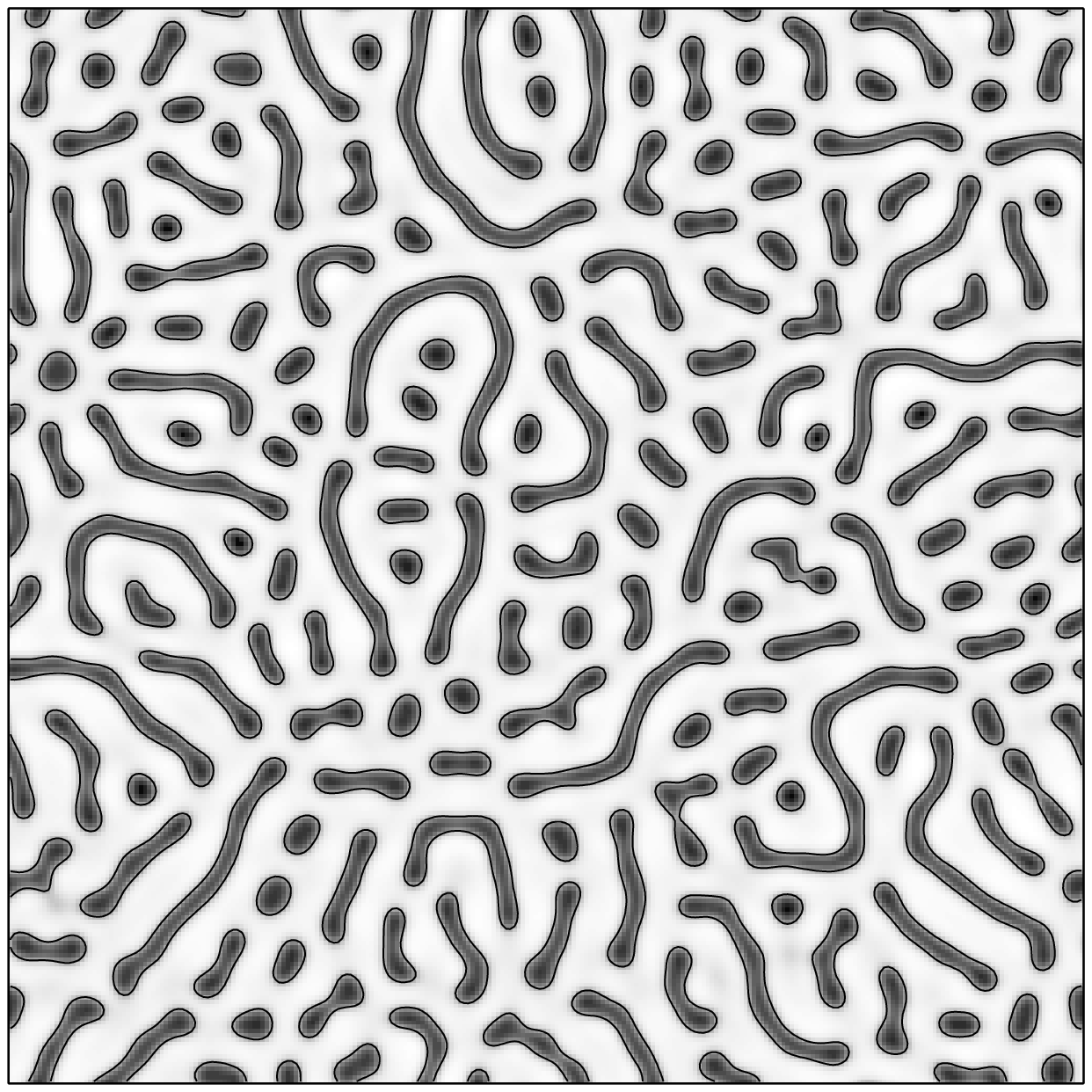,width=40mm} \\
\epsfig{file=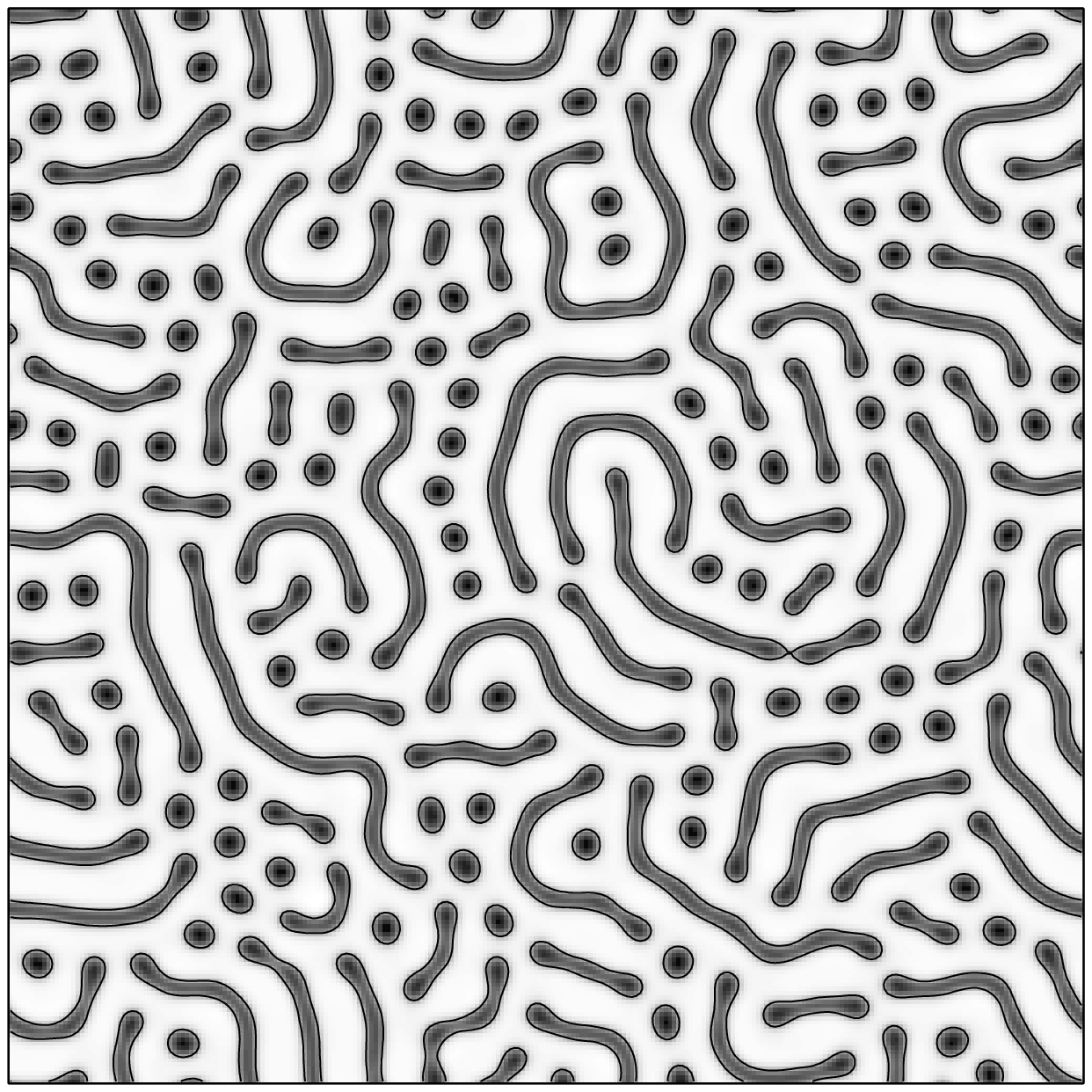,width=40mm} & \epsfig{file=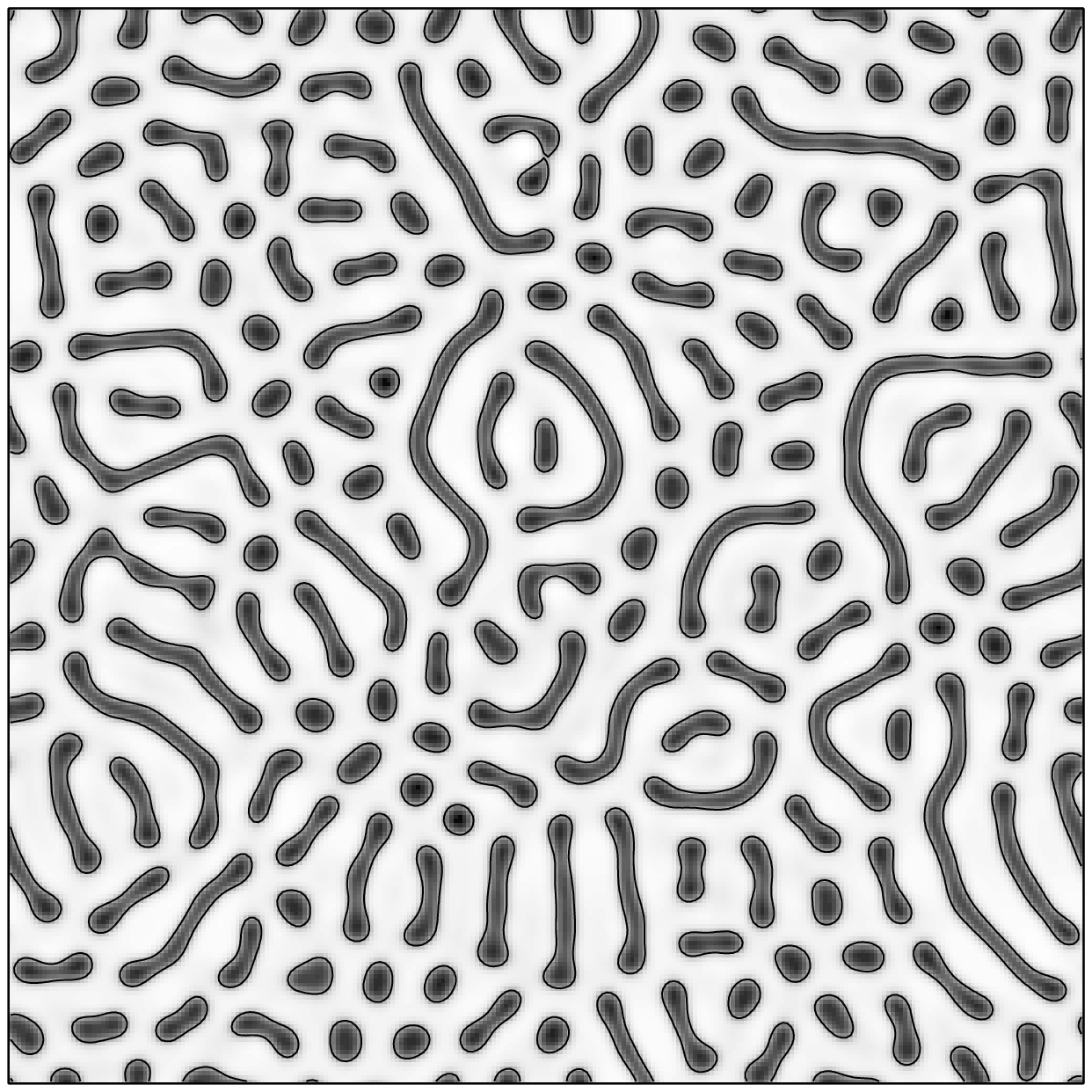,width=40mm}
\end{tabular} 
\caption{ Time evolution of the domain structure for a quadratic reaction mechanism with rate $\Gamma=0.001$ for high viscosity ($\tau_{f}=400$, left-hand column) and low viscosity ($\tau_{f}=5.0$, right-hand column) at times $t=10,000$ (top), $t=40,000$ (middle) and $t=200,000$ (bottom). The $A$-rich regions are shown in white and the $B$-rich regions in black.}
\label{fig:qRfastRate}
\end{figure}
shows the evolution of the domain structure at high viscosity for $\Gamma=0.0001$. At early times there is a period of domain growth. However, due to the asymmetry of the reaction mechanism, the growth of the $B$-rich domains is halted at a certain size after which the $B$-rich domains are depleted by the reaction with smaller domains vanishing completely. As this process occurs some of the larger domains break-up to reduce the surface tension. This creates small, circular domains which either vanish or assume a steady size. At some point a balance is reached between the depletion of the $B$-rich domains and their growth due to B, produced in the surrounding $A$-rich fluid, being supplied by diffusion.

If the viscosity is decreased the initial rate of growth is higher as shown in the right-hand column of Figure~\ref{fig:qRslowRate}. As at high viscosity the growth of $B$-rich domains is eventually reversed by the reaction. However, as the $B$-rich domains are depleted hydrodynamic flows continue to drive coalescence of nearby domains faster than is possible by diffusion alone. Consequently, at intermediate times, there are fewer domains and they are less elongated than at high viscosity. At later times, gradients of $\phi$ within the $A$-matrix trigger several secondary phase separations, increasing the density of domains at the latest times simulated. By $t\sim 200,000$ an approximately steady state has been reached.

For a faster rate of reaction, $\Gamma=0.001$, `worm-like' phases are able to persist until the latest time simulated as shown in the left-hand column of Figure~\ref{fig:qRfastRate}.
 This indicates a transition to a reaction-dominated regime where structures with high curvature, which are unfavoured by flow and diffusion, can persist for a long time. 

If the viscosity is decreased, as shown on the right of Figure~\ref{fig:qRfastRate}, the structures formed are similar to those at high viscosity. However, secondary phase separations are triggered within the A-rich phase increasing the number of small domains. 

\begin{figure}
\psfrag{L}{$l_{I}$}
\psfrag{t}{$t$}
\centering
\begin{tabular}{ccccc}
\epsfig{file=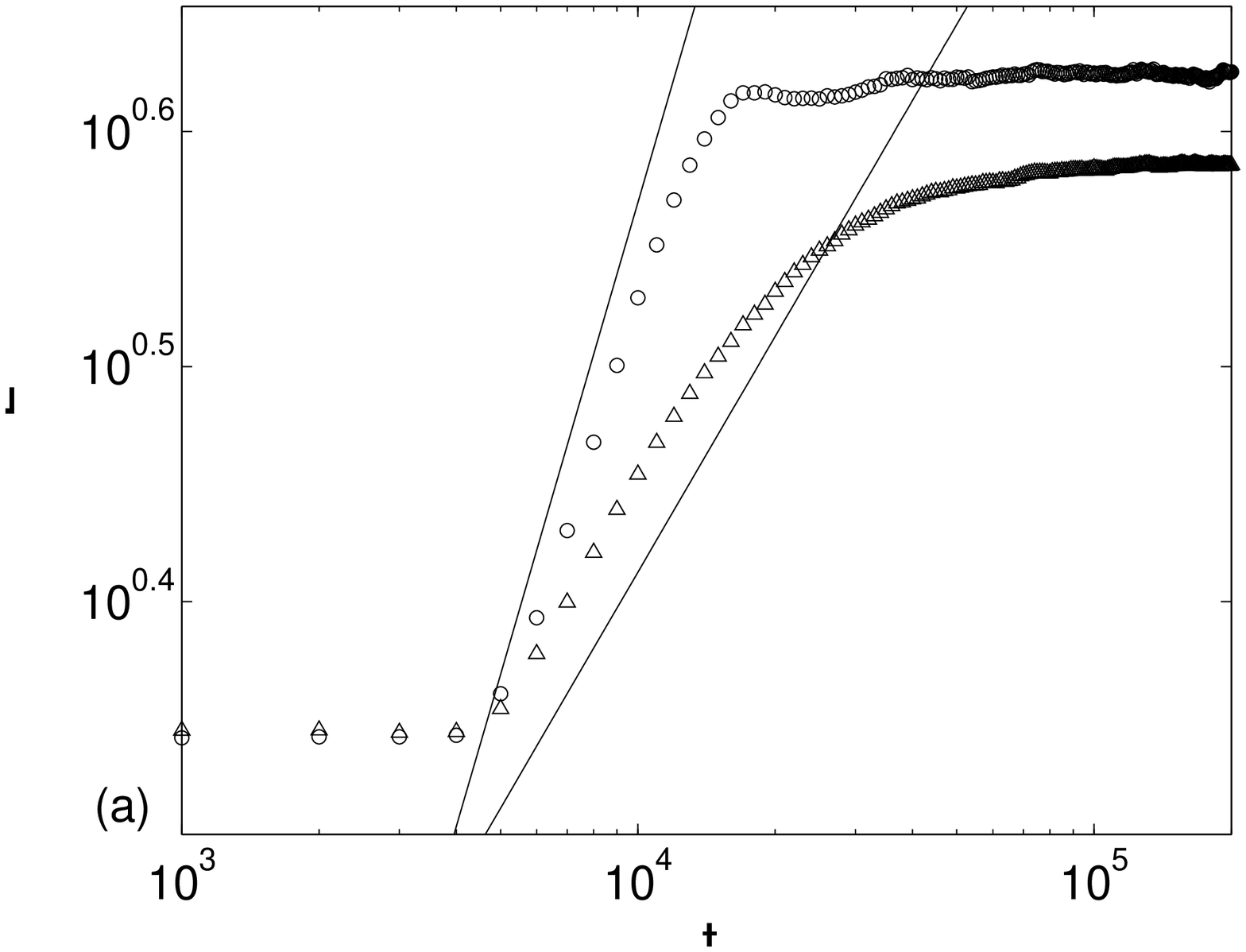,width=60mm} & & & & \epsfig{file=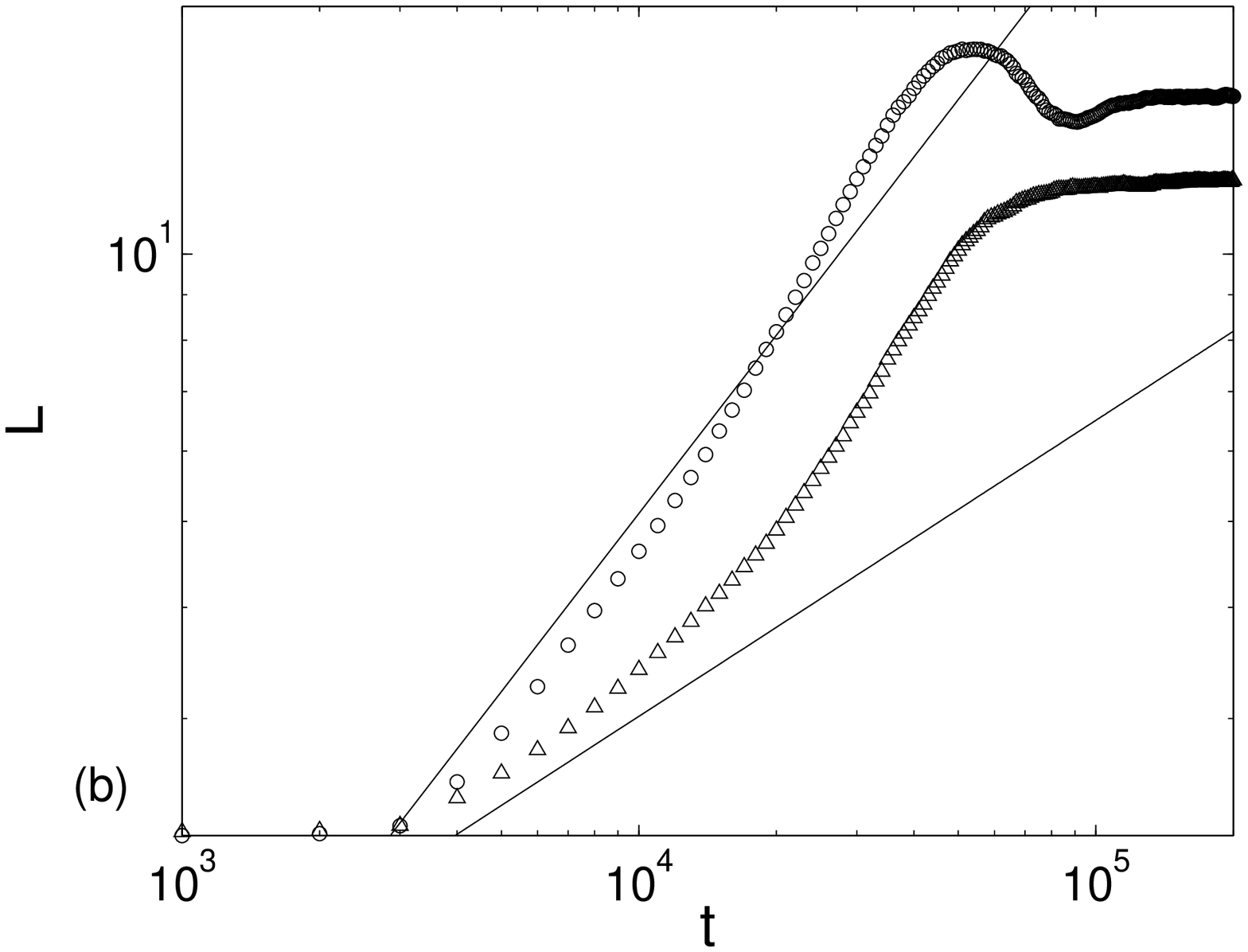,width=60mm}
\end{tabular}
\caption{ Log-log plot of the average domain size measured by the inverse interfacial length for (a)~$\Gamma=0.001$ and (b)~$\Gamma=0.0001$ and values of $\tau_f=400$ (high viscosity, $\triangle$) and $\tau_{f}=5.0$ (low viscosity, $\circ$). The solid straight lines correspond to $\alpha=1/3$ and $\alpha=2/3$.}
\label{fig:qRScaling}
\end{figure}

Figure~\ref{fig:qRScaling} shows
the evolution of the inverse interfacial length for different values of $\Gamma$ and $\nu$. It can be seen that when hydrodynamic effects are present the rate of reduction of interfacial length is faster. Moreover, $l_{I}$ is no longer monotonically increasing: it attains a maximum due to the creation of more interface during secondary phase separation events.
\section{Conclusions}
In conclusion, we have extended the lattice Boltzmann method to study the effect of hydrodynamics on structures arising in phase-separating reactive mixtures for two simple reaction mechanisms. We have found that hydrodynamic flow significantly alters both the way in which the domain structure in these fluids evolves and the eventual steady states of the system. The results obtained for the linear reaction are in agreement with previous work on the subject. For a quadratic reaction an asymmetric domain structure was obtained, with the inclusion of hydrodynamic effects leading to secondary phase separation within majority phase.

It would be interesting to extend the model to incorporate viscoelastic effects arising from constituent molecules  with internal microstructure: one can envisage chemical processes which can change the local microstructural elements. The technique also makes feasible a study of the interaction of reaction and diffusion with imposed flow-fields.

\section{Acknowledgements}\nonumber
We would like to thank C.~Pooley for intersting discussions. K.F. acknowledges funding from EPRSC Grant No.~GR/R83712/01.


\begin{thebibliography}{99}

\bibitem{glotzer} S.C. Glotzer, E.A. Di~Marzio and M. Muthukumar, Phys. Rev. Lett. {\bf 74} 2034 (1995). 
\bibitem{christen} J.J. Christensen, K. Elder and H.C. Fogedby, Phys. Rev. E {\bf 54} R2212 (1996).
\bibitem{glotzer2} S.C. Glotzer and A. Coniglio, Phys. Rev. E {\bf 50} 4241 (1994).
\bibitem{leiber} L. Leiber, Marcomolecules {\bf 13} 1602 (1980).
\bibitem{ohta} T. Ohta and K. Kawasaki, Marcomolecules {\bf 19} 2621 (1986).
\bibitem{okuz} T. Okuzono and T. Ohta, Phys. Rev. E {\bf 67} 056211 (2003).
\bibitem{liu}  B. Liu, C. Tong and Y. Yang, J. Phys. Chem. B {\bf 105} 10091 (2001).
\bibitem{tong} C. Tong, H. Zhang and Y. Yang,  J. Phys. Chem. B {\bf 106} 7869 (2002).
\bibitem{balazs} R.D.M. Travasso, G.A. Buxton, O. Kuksenok, K. Good and A.C. Balazs, J. Chem. Phys. {\bf 122} 194906 (2005).
\bibitem{reig} R. Reigada, F. Sagues and A.S. Mikhailov, Phys. Rev. Lett. {\bf 89} 038301 (2002); R. Reigada, A.S. Mikhailov and F. Sagues, Phys. Rev. E {\bf 69} 041103 (2004).
\bibitem{wagner} A.J. Wagner and J.M. Yeomans, Phys. Rev. Lett. {\bf 80} 1429 (1998).
\bibitem{huo} Y. Huo, X. Jiang and Y. Yang, J. Chem. Phys. {\bf 118} 9830 (2003); Macromol. Theory Simul. {\bf 13} 280 (2004).
\bibitem{photod} Q. Tran-Cong and A. Harada, Phys. Rev. Lett. {\bf 76} 1162 (1996); Q. Tran-Cong, A. Harada, K. Kataoka, T. Ohta and O. Urakawa,  Phys. Rev. E  {\bf 55} R6340 (1997); Qui Tran-Cong, Katsunari Kataoka and Osamu Urakawa, Phys. Rev. E {\bf 57} R1243 (1998).
\bibitem{photoi} Qui Tran-Cong, Takashi Ohta and Osamu Urakawa, Phys. Rev. E {\bf 56} R59 (1997); T. Ohta, O. Urakawa and Q. Tran-Cong, Macromolecules {\bf 31} 6845 (1998).
\bibitem{tra} Q. Tran-Cong, J. Kawai and K. Endoh, Chaos {\bf 9} 298 (1991).
\bibitem{kendon} V.M. Kendon, M.E. Cates, I. Pagonanarraga, J-C. Desplat and P. Bladon, J. Fluid. Mech. {\bf 440} 147 (2001).
\bibitem{bray} A.J. Bray, Adv. Phys. {\bf 43} 357 (1994).
\bibitem{siggia} E.D. Siggia, Phys. Rev. A {\bf 20} 595 (1979).
\bibitem{tan} H. Tanaka, Phys. Rev. E {\bf 51} 1313 (1995).
\bibitem{cahn} J.W. Cahn, J. Chem. Phys. {\bf 42} 93 (1964).
\bibitem{succi} S. Succi, {\it The Lattice Boltzmann Equation for Fluid Dynamics and Beyond} (Clarendon Press, Oxford, 2001).
\bibitem{shan1} Xiaowen Shan and Hudong Chen, Phys. Rev. E {\bf 47} 1815 (1993).
\bibitem{swift} M.R. Swift, E. Orlandini, W.R. Osborn and J.M. Yeomans, Phys. Rev. E {\bf 54} 5041 (1996).
\bibitem{giri} L-S. Luo and S.S. Girimaji, Phys. Rev. E {\bf 67} 036302 (2003).
\bibitem{shan2} Xiaowen Shan and Hudong Chen, Phys. Rev. E {\bf 49} 2941 (1994).
\bibitem{luo} Li-Shi Luo, Phys. Rev. Lett. {\bf 81} 1618 (1998).
\bibitem{dawson} S. Pounce Dawson, S. Chen and G.D. Doolen, J. Chem. Phys. {\bf 98} 1514 (1992).
\bibitem{denn} C. Denniston, E. Orlandini and J.M. Yeomans, Europhys. Lett. {\bf 52} 481 (2000).
\bibitem{ladd} A.J.C. Ladd and R. Verberg, J. Stat. Phys. {\bf 104} 1191 (2001). 

\end{thebibliography}
\end{document}